\PassOptionsToPackage{unicode}{hyperref}
\PassOptionsToPackage{hyphens}{url}
\documentclass[
  11pt,
]{article}
\usepackage{xcolor}
\usepackage{amsmath,amssymb}
\usepackage{iftex}
\ifPDFTeX
  \usepackage[T1]{fontenc}
  \usepackage[utf8]{inputenc}
  \usepackage{textcomp} 
\else 
  \usepackage{unicode-math} 
  \defaultfontfeatures{Scale=MatchLowercase}
  \defaultfontfeatures[\rmfamily]{Ligatures=TeX,Scale=1}
\fi
\usepackage{lmodern}
\ifPDFTeX\else
\fi
\IfFileExists{upquote.sty}{\usepackage{upquote}}{}
\IfFileExists{microtype.sty}{
  \usepackage[]{microtype}
  \UseMicrotypeSet[protrusion]{basicmath} 
}{}
\makeatletter
\@ifundefined{KOMAClassName}{
  \IfFileExists{parskip.sty}{%
    \usepackage{parskip}
  }{
    \setlength{\parindent}{0pt}
    \setlength{\parskip}{6pt plus 2pt minus 1pt}}
}{
  \KOMAoptions{parskip=half}}
\makeatother
\usepackage{color}
\usepackage{fancyvrb}

\DefineVerbatimEnvironment{Highlighting}{Verbatim}{commandchars=\\\{\}}
\newenvironment{Shaded}{}{}

\newcommand{\AttributeTok}[1]{\textcolor[rgb]{0.49,0.56,0.16}{#1}}

\newcommand{\BuiltInTok}[1]{\textcolor[rgb]{0.00,0.50,0.00}{#1}}

\newcommand{\CommentTok}[1]{\textcolor[rgb]{0.38,0.63,0.69}{\textit{#1}}}

\newcommand{\ControlFlowTok}[1]{\textcolor[rgb]{0.00,0.44,0.13}{\textbf{#1}}}

\newcommand{\DecValTok}[1]{\textcolor[rgb]{0.25,0.63,0.44}{#1}}

\newcommand{\ImportTok}[1]{\textcolor[rgb]{0.00,0.50,0.00}{\textbf{#1}}}

\newcommand{\KeywordTok}[1]{\textcolor[rgb]{0.00,0.44,0.13}{\textbf{#1}}}
\newcommand{\NormalTok}[1]{#1}
\newcommand{\OperatorTok}[1]{\textcolor[rgb]{0.40,0.40,0.40}{#1}}

\newcommand{\StringTok}[1]{\textcolor[rgb]{0.25,0.44,0.63}{#1}}
\newcommand{\VariableTok}[1]{\textcolor[rgb]{0.10,0.09,0.49}{#1}}

\usepackage{longtable,booktabs,array}
\usepackage{calc} 
\usepackage{etoolbox}
\makeatletter
\patchcmd\longtable{\par}{\if@noskipsec\mbox{}\fi\par}{}{}
\makeatother
\IfFileExists{footnotehyper.sty}{\usepackage{footnotehyper}}{\usepackage{footnote}}
\makesavenoteenv{longtable}
\usepackage{graphicx}
\makeatletter
\newsavebox\pandoc@box
\newcommand*\pandocbounded[1]{
  \sbox\pandoc@box{#1}%
  \Gscale@div\@tempa{\textheight}{\dimexpr\ht\pandoc@box+\dp\pandoc@box\relax}%
  \Gscale@div\@tempb{\linewidth}{\wd\pandoc@box}%
  \ifdim\@tempb\p@<\@tempa\p@\let\@tempa\@tempb\fi
  \ifdim\@tempa\p@<\p@\scalebox{\@tempa}{\usebox\pandoc@box}%
  \else\usebox{\pandoc@box}%
  \fi%
}
\def\fps@figure{htbp}
\makeatother
\providecommand{\tightlist}{%
  \setlength{\itemsep}{0pt}\setlength{\parskip}{0pt}}
\usepackage[margin=1in]{geometry}
\usepackage{amsmath,amssymb,amsthm}
\usepackage{graphicx}
\usepackage{booktabs}
\usepackage{longtable}
\usepackage{array}
\usepackage{float}
\usepackage{xcolor}
\usepackage{listings}
\usepackage{fancyvrb}
\usepackage[hidelinks]{hyperref}
\usepackage{microtype}
\usepackage{enumitem}
\setlist{topsep=3pt,itemsep=2pt,parsep=0pt}

\usepackage{newunicodechar}
\newunicodechar{→}{\ensuremath{\rightarrow}}
\newunicodechar{←}{\ensuremath{\leftarrow}}
\newunicodechar{↑}{\ensuremath{\uparrow}}
\newunicodechar{↓}{\ensuremath{\downarrow}}
\newunicodechar{≥}{\ensuremath{\geq}}
\newunicodechar{≤}{\ensuremath{\leq}}
\newunicodechar{≈}{\ensuremath{\approx}}
\newunicodechar{≠}{\ensuremath{\neq}}
\newunicodechar{∧}{\ensuremath{\wedge}}

\usepackage{bookmark}
\IfFileExists{xurl.sty}{\usepackage{xurl}}{} 
\hypersetup{
  pdftitle={Cost-Aware Speculative Execution for LLM-Agent Workflows: An Integrated Five-Dimension Method},
  pdfauthor={Faisal Fareed, AWS},
  hidelinks,
  pdfcreator={LaTeX via pandoc}}

\title{Cost-Aware Speculative Execution for LLM-Agent Workflows: An
Integrated Five-Dimension Method}
\author{Faisal Fareed, AWS}
\date{June 5, 2026}

\begin{document}
\maketitle

\subsection{Abstract}\label{abstract}

LLM-agent workflows chain large-language-model calls and tool
invocations, and they spend most of their wall-clock time waiting on
upstream operations before downstream ones can start. Speculative
execution can reclaim that idle time by launching a downstream operation
with a predicted upstream input, but in this setting each speculation
costs real money (per-token API billing) and the success probability is
hard to estimate and changes over time. This paper describes a method
organized around five design decisions: (D1) start a downstream
operation before its upstream completes, (D2) price each speculation in
real dollars at separate input and output rates, (D3) expose a single
operator dial for latency versus cost, (D4) decide using an
expected-value rule with a failure-weighted cost term and a
preference-adjusted threshold, and (D5) estimate the success probability
with a Bayesian Beta-Binomial posterior whose prior is keyed to a
dependency-type taxonomy. Variants of these ideas appear in recent work;
the combination, with every decision logged in dollars, is what is new.
The decision rule fires only on edges that pass an admissibility
precondition (the speculated downstream must be side-effect-free,
idempotent, or stageable behind a commit barrier), because a wrong
speculation is rolled back by re-execution, which refunds wasted tokens
but cannot un-send an irreversible side effect. The paper also specifies
the runtime mechanics: a two-phase plan-plus-runtime decision model with
bidirectional override, and streaming re-estimation of the predicted
input with mid-stream cancellation and fractional-waste accounting. It
gives a closed-form derivation that the decision rule self-limits as the
upstream branching factor grows. For deployment it specifies a
five-stage calibration pipeline (offline replay → shadow → canary with
recovery of the implied dollar-value-of-latency → online calibration →
drift-triggered kill-switch) and a workload-fit rubric covering eight
production archetypes. Contrast tables against the four closest
published systems, namely Dynamic Speculative Agent Planning (DSP),
Speculative Actions v2, Sherlock, and B-PASTE, show clear
differentiators on every dimension. A synthetic numerical validation
suite (Appendix D) confirms, at the canonical AutoReply parameters, that
the speculate-or-wait boundary matches the closed-form critical
branching factor, that the expected value crosses zero at the predicted
success-probability threshold, that the Beta-Binomial posterior recovers
the true success rate from a uniform prior, that streaming cancellation
reduces per-failure waste as predicted, and that the implied-value audit
catches the divergent operating point used as the running example.

\textbf{Keywords:} speculative execution, LLM agents, agent workflows,
cost-aware scheduling, Bayesian inference, dependency types, multi-agent
orchestration.

\subsection{1. Introduction}\label{introduction}

\subsubsection{1.1 Setting}\label{setting}

An LLM-agent workflow is a chain of steps in which each step feeds the
next --- formally, a directed acyclic graph (DAG) of operations. Each
step (vertex) is an LLM call or a tool invocation, and each dependency
(edge) means one step consumes another's output. Two things make these
workflows expensive. First, they are slow: LLM latency dominates
execution time, with a single operation typically taking hundreds of
milliseconds to tens of seconds, and end-to-end latency is the
critical-path sum of those operation latencies minus whatever
parallelism the runtime extracts. Second, they are costly: cost is the
sum of per-token billing, which for major commercial APIs is asymmetric
(output tokens billed at 3--8× the input-token rate) and varies
substantially across providers and model tiers.

Speculative execution attacks the latency by starting a downstream step
before its upstream has finished, working from a best-guess input in
place of the output the upstream will eventually produce. Formally, it
starts a downstream operation \texttt{v} before its upstream \texttt{u}
has produced output, using a predicted input \texttt{î} in place of
\texttt{u}'s eventual output \texttt{i}. When the guess is good enough
--- the downstream result computed from \texttt{î} is acceptable given
\texttt{i} --- end-to-end wall-clock time decreases. When it is not, the
downstream is re-executed with \texttt{i} and the speculative cost is
wasted.

\subsubsection{1.2 The research question}\label{the-research-question}

\begin{quote}
When should an LLM-agent runtime launch a downstream operation before
its upstream has finished, given that each speculation costs real money,
the chance the guess pays off is hard to estimate and drifts over time,
and operators need to trade speed against cost differently at different
times?
\end{quote}

\subsubsection{1.3 Contributions}\label{contributions}

The method has five design dimensions and seven auxiliary mechanisms:

\begin{enumerate}
\def\labelenumi{\arabic{enumi}.}
\tightlist
\item
  \textbf{D1.} Pre-upstream-completion speculation over arbitrary DAGs
  of LLM operations (Section 3).
\item
  \textbf{D2.} Two-rate per-token monetary cost, plugged into the
  decision rule at runtime (Section 4).
\item
  \textbf{D3.} A user-facing \texttt{α\ ∈\ {[}0,1{]}} preference dial,
  separated from a deployment-configured \texttt{λ} (\$/second)
  latency-value conversion (Section 5).
\item
  \textbf{D4.} Expected-value decision rule with failure-weighted waste
  term and \texttt{α}-scaled threshold (Section 6).
\item
  \textbf{D5.} Bayesian Beta-Binomial posterior over success
  probability, with structural priors keyed to an LLM dependency-type
  taxonomy (Section 7).
\item
  Two-phase plan-time + runtime decision model with bidirectional
  override (Section 8).
\item
  Streaming partial-output re-estimation with mid-stream cancellation
  and fractional waste accounting (Section 9).
\item
  Self-limiting behavior under branching factor with closed-form
  critical-\texttt{k} derivation and effective-\texttt{k} accounting
  under skew (Section 7.6).
\item
  Five-stage calibration-and-evaluation pipeline, including
  implied-\texttt{λ} recovery from observed operator preferences
  (Section 12).
\item
  Four-point workload-fit rubric with eight production archetypes and
  four explicit non-fit shapes (Section 13).
\item
  Per-decision telemetry schema sufficient for every calibration signal
  (Appendix C).
\item
  Synthetic numerical validation suite covering decision boundary,
  P-threshold, posterior convergence, streaming cancellation, and
  implied-\texttt{λ} recovery, each fully specified from the equations
  under a single fixed seed (Appendix D).
\end{enumerate}

Each item is a specific design choice with a real alternative the paper
rejects. Section 11 shows that no subset of the audited prior art (DSP,
Speculative Actions v2, Sherlock, B-PASTE) discloses the combination.
Appendices A--D give the posterior-update mechanics, a worked
router-dependency example, the per-decision telemetry schema, and a
synthetic numerical validation suite.

\subsubsection{1.4 Scope}\label{scope}

\textbf{In scope.} Cost-aware speculative execution for LLM-agent
workflows expressed as a DAG of LLM calls and tool invocations, billed
per token or per GPU-hour.

\textbf{Out of scope.} Token-level speculative decoding inside a single
LLM forward pass (a different layer of the stack); dynamic workflows
whose topology is determined at runtime (loops, reflection, agent
spawning); RLHF and alignment questions; settings where the runtime
cannot produce a predicted input \texttt{î} at the moment a speculation
could fire; and downstream operations whose external side effects are
not side-effect-free, idempotent, or stageable behind a commit barrier
(the admissibility precondition of Section 3.3, since a speculation that
sends an email or charges a card cannot be rolled back by re-execution).

\subsection{2. Problem and notation}\label{problem-and-notation}

\subsubsection{2.1 Setting}\label{setting-1}

An LLM-agent workflow is a DAG \texttt{W\ =\ (V,\ E)} where each vertex
\texttt{v\ ∈\ V} is an LLM call or tool invocation and each edge
\texttt{(u,\ v)\ ∈\ E} indicates that \texttt{v} consumes output from
\texttt{u}. A speculative-execution runtime decides, for each downstream
operation \texttt{v} whose upstream \texttt{u} is currently executing,
whether to launch \texttt{v} before \texttt{u} completes (using a
predicted upstream input \texttt{î}) or wait for \texttt{u} to finish.

\pandocbounded{\includegraphics[keepaspectratio]{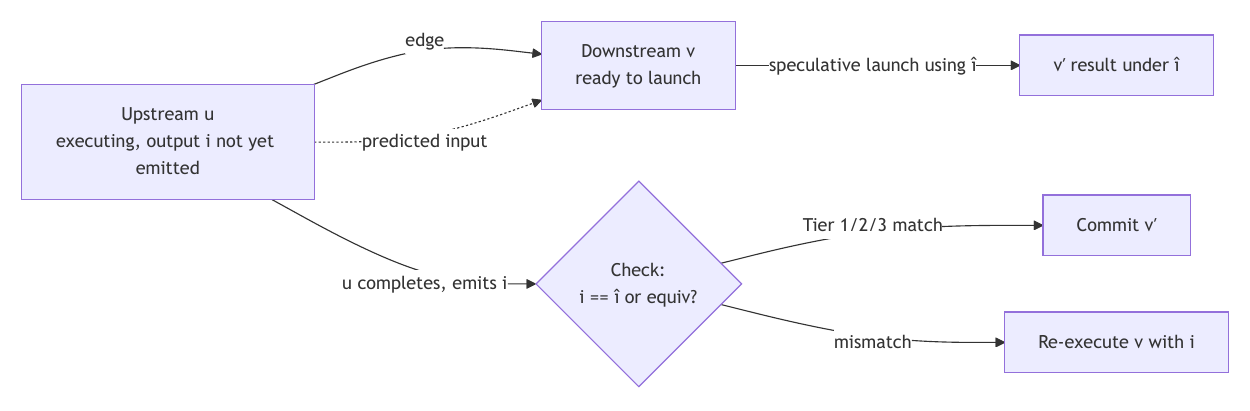}}

\emph{Figure 1. DAG speculation geometry. The runtime launches
\texttt{v} against a predicted input \texttt{î} while \texttt{u} is
still generating. On \texttt{u}'s completion, a three-tier check
(Section 7.4) decides whether \texttt{v}'s speculative result is
reusable.}

\subsubsection{2.2 Notation}\label{notation}

{\def\LTcaptype{none} 
\begin{longtable}[]{@{}
  >{\raggedright\arraybackslash}p{(\linewidth - 4\tabcolsep) * \real{0.3333}}
  >{\raggedright\arraybackslash}p{(\linewidth - 4\tabcolsep) * \real{0.3333}}
  >{\raggedright\arraybackslash}p{(\linewidth - 4\tabcolsep) * \real{0.3333}}@{}}
\toprule\noalign{}
\begin{minipage}[b]{\linewidth}\raggedright
Symbol
\end{minipage} & \begin{minipage}[b]{\linewidth}\raggedright
Meaning
\end{minipage} & \begin{minipage}[b]{\linewidth}\raggedright
Units
\end{minipage} \\
\midrule\noalign{}
\endhead
\bottomrule\noalign{}
\endlastfoot
\texttt{P} & probability that a speculation is useful & dimensionless,
\texttt{{[}0,1{]}} \\
\texttt{C\_spec} & monetary cost of speculatively executing the
downstream & USD \\
\texttt{L\_value} & dollar-value of latency saved on speculation success
& USD \\
\texttt{L} & estimated latency savings & seconds \\
\texttt{α} & user preference (0 = cost-sensitive, 1 = latency-sensitive)
& dimensionless, \texttt{{[}0,1{]}} \\
\texttt{λ} & latency value ratio (dollars per second saved) & USD/s \\
\texttt{input\_tokens}, \texttt{output\_tokens} & estimated token counts
for the speculative operation & tokens \\
\texttt{input\_price}, \texttt{output\_price} & API billing rates &
USD/token \\
\texttt{î}, \texttt{i} & predicted upstream input, actual upstream input
& n/a \\
\end{longtable}
}

By convention all money is in US dollars; currency conversion is a
deployment detail.

\subsection{3. Dimension 1: Pre-upstream-completion
speculation}\label{dimension-1-pre-upstream-completion-speculation}

\subsubsection{3.1 Definition}\label{definition}

For a downstream operation \texttt{v} that depends on an upstream
\texttt{u}, the runtime may launch \texttt{v} \textbf{before \texttt{u}
has produced its output}, using a predicted input \texttt{î} in place of
the real input \texttt{i} that \texttt{u} will eventually emit. When
\texttt{u} completes and emits \texttt{i}, the runtime checks whether
\texttt{v}'s speculative result (computed using \texttt{î}) is
acceptable per the three-tier criterion of Section 7.4. If acceptable,
the speculative result is committed; if not, \texttt{v} is re-executed
with \texttt{i}.

This is mechanically distinct from three adjacent patterns:

\begin{itemize}
\tightlist
\item
  \textbf{Token-level speculative decoding} {[}Leviathan et al., 2023{]}
  operates within a single LLM forward pass, not across agent
  operations.
\item
  \textbf{Post-output, pre-verification speculation} {[}Ro et al.,
  2025{]} uses the actual upstream output \texttt{i} and speculates that
  \emph{verification} of \texttt{i} will pass.
\item
  \textbf{Reasoning-time tool pre-launch} {[}Sui et al., 2026; Song,
  2026{]} speculates the next tool call while the agent LLM is still
  generating its next plan step.
\end{itemize}

D1 is specifically \emph{``upstream has not yet produced \texttt{i}; we
guess \texttt{î} now.''}

\subsubsection{3.2 Where the predicted input comes
from}\label{where-the-predicted-input-comes-from}

Three sources, in preference order:

\begin{enumerate}
\def\labelenumi{\arabic{enumi}.}
\tightlist
\item
  \textbf{Context-conditioned prediction.} A cheap auxiliary model or
  template predicts \texttt{î} from the upstream's input and partial
  state. If the upstream is a topic-extraction agent known to produce
  3--5 topics, \texttt{î} may be \emph{``the top-ranked candidate topic
  from the upstream's partial state.''}
\item
  \textbf{Most-likely historical input.} From logged
  \texttt{(upstream\_input,\ upstream\_output)} pairs, the modal output
  for similar inputs.
\item
  \textbf{Streaming partial output} (Section 9). If the upstream streams
  tokens, re-estimate \texttt{î} as tokens arrive, updating the
  speculation decision in flight.
\end{enumerate}

\texttt{î} is a design choice; the \emph{correctness} of the rest of the
method does not depend on \emph{how} \texttt{î} was produced, only that
(a) there is a predicted input at speculation-launch time and (b) the
success criterion of Section 7.4 labels each trial. The method's latency
economics do depend on the predictor's own cost: a predictor that is
itself an expensive model call can erode or erase the net latency saving
(Section 14.2).

\subsubsection{3.3 Admissibility: speculation requires a
side-effect-free or compensable
downstream}\label{admissibility-speculation-requires-a-side-effect-free-or-compensable-downstream}

A speculation may turn out to be wrong: when \texttt{u} emits \texttt{i}
and the tier-1/2 check (Section 7.4) fails, the method re-executes
\texttt{v} with the correct input. This rollback is sound only if
launching \texttt{v} against the wrong \texttt{î} left no observable
trace outside the runtime. The cost model (Section 4) accounts for the
\emph{wasted tokens} of a failed speculation, but it does not account
for a failed speculation's \emph{external effects}. Re-execution refunds
neither.

A downstream operation \texttt{v} is therefore \textbf{admissible for
speculation} only if at least one of the following holds:

\begin{enumerate}
\def\labelenumi{\arabic{enumi}.}
\tightlist
\item
  \textbf{Side-effect-free.} \texttt{v} is a pure LLM generation or a
  read-only tool call (retrieval, a \texttt{GET}, a lookup). Discarding
  its result on tier-failure costs only tokens. This is the default and
  covers every archetype in Section 13.
\item
  \textbf{Idempotent under the natural key.} \texttt{v}'s effect is
  keyed such that the speculative invocation and the corrected
  re-execution collapse to the same final state (e.g., an upsert keyed
  on a deterministic id). The speculative write is overwritten, not
  duplicated.
\item
  \textbf{Staged behind a commit barrier.} \texttt{v}'s
  externally-visible effect is buffered (a draft, a transaction not yet
  committed, an outbound message held in a queue) and released only
  after the tier-1/2 check passes. On failure the staged effect is
  dropped and \texttt{v} re-runs before anything is released.
\end{enumerate}

Operations that fail all three (a tool call that sends an email, charges
a card, posts an irreversible mutation, or triggers a downstream
actuator the moment it is invoked) \textbf{must not be speculated},
regardless of EV. Re-execution cannot un-send the message, and
\texttt{(1−P)\ ·\ C\_spec} does not price the cost of the erroneous side
effect. Such edges are tagged \texttt{non\_speculable} and the per-edge
enable bit (Section 12) is held off for them at deployment time,
independent of the decision rule. The EV gate of Section 6 runs only on
edges that have already passed this admissibility test.

This is a hard precondition, not a tuning knob: the failure-weighted
cost model is correct precisely because the only thing wasted on a
failed speculation is compute. That assumption holds exactly when
\texttt{v} is admissible by one of the three routes above.

\subsection{4. Dimension 2: Per-token monetary cost in
decisions}\label{dimension-2-per-token-monetary-cost-in-decisions}

\subsubsection{4.1 Two-rate per-token
pricing}\label{two-rate-per-token-pricing}

For an LLM operation billed per token at distinct input and output
rates,

\begin{verbatim}
C_spec = input_tokens · input_price + output_tokens · output_price .
\end{verbatim}

\texttt{input\_tokens} and \texttt{output\_tokens} are estimates
available at speculation time. Commercial APIs (Anthropic, OpenAI,
Google, Mistral) bill input and output tokens at distinct rates
differing by 3--8× as of 2026, so conflating them into a single rate
materially distorts the decision at high output-to-input ratios (the
common case for generation-heavy agents).

A pricing-map data structure records per-\texttt{(provider,\ model)}
rates:

\begin{Shaded}
\begin{Highlighting}[]
\AttributeTok{@dataclass}
\KeywordTok{class}\NormalTok{ PricingEntry:}
\NormalTok{    provider: }\BuiltInTok{str}                       \CommentTok{\# e.g. "anthropic", "openai"}
\NormalTok{    model: }\BuiltInTok{str}                          \CommentTok{\# e.g. "claude{-}opus{-}4{-}7"}
\NormalTok{    input\_price\_per\_token: }\BuiltInTok{float}        \CommentTok{\# USD per input token}
\NormalTok{    output\_price\_per\_token: }\BuiltInTok{float}       \CommentTok{\# USD per output token}

\NormalTok{PRICING\_MAP: }\BuiltInTok{dict}\NormalTok{[}\BuiltInTok{tuple}\NormalTok{[}\BuiltInTok{str}\NormalTok{, }\BuiltInTok{str}\NormalTok{], PricingEntry] }\OperatorTok{=}\NormalTok{ \{...\}}
\end{Highlighting}
\end{Shaded}

\subsubsection{4.2 Token estimation}\label{token-estimation}

The speculative input \texttt{î} is known by construction of D1, so
\texttt{input\_tokens\ =\ tokenize(prompt(î)).length} is computable
exactly. Output-token estimation is harder because LLM output length is
non-deterministic. Three acceptable approaches:

\begin{itemize}
\tightlist
\item
  \textbf{EMA over historical output lengths} for the same
  \texttt{(agent,\ input-shape)} pair with decay constant
  \texttt{α\_EMA\ =\ 0.2} (default).
\item
  \textbf{Fixed-ceiling policy} with
  \texttt{max\_tokens\ =\ estimated\ +\ 2σ}.
\item
  \textbf{Deployment override} where operators supply a conservative
  point estimate per agent.
\end{itemize}

The method is robust to moderate estimation error (Section 10.2
sensitivity analysis), but the decision rule's sensitivity to
\texttt{C\_spec} grows as \texttt{P} decreases, so low-\texttt{P} /
high-variance agents should be tagged \texttt{uncertain\_cost} and
excluded from speculation until history stabilizes.

\subsubsection{4.3 Other cost models}\label{other-cost-models}

\texttt{C\_spec} is intentionally pluggable. For self-hosted open-source
models, \texttt{C\_spec} may be computed via GPU-hour amortization:

\begin{verbatim}
C_spec = (unit_price · num_gpus · output_tokens) / (throughput · utilization) .
\end{verbatim}

This reduces to a linear-per-token form, so the decision rule is
unchanged. The \textbf{distinctive} choice is the two-rate (input ≠
output) form at API-billing granularity; single-rate GPU-hour forms do
not fully exploit the billing asymmetry.

\subsection{\texorpdfstring{5. Dimension 3: User-facing \texttt{α}
preference
dial}{5. Dimension 3: User-facing α preference dial}}\label{dimension-3-user-facing-ux3b1-preference-dial}

\subsubsection{5.1 Semantics}\label{semantics}

\texttt{α} is a dimensionless scalar on \texttt{{[}0,1{]}} exposed to
the operator (or end user) that expresses the relative preference
between minimizing latency and minimizing cost.

\begin{itemize}
\tightlist
\item
  \texttt{α\ =\ 0}: fully cost-sensitive. Speculate only when
  \texttt{EV\ ≥\ C\_spec} (full-cost break-even).
\item
  \texttt{α\ =\ 1}: fully latency-sensitive. Speculate whenever
  \texttt{EV\ ≥\ 0}.
\item
  \texttt{α\ =\ 0.5}: balanced.
\end{itemize}

\texttt{α} drives the decision threshold via
\texttt{threshold\ =\ (1−α)\ ·\ C\_spec} (Section 6).

\subsubsection{5.2 Runtime mutability}\label{runtime-mutability}

\texttt{α} may be changed at any time during workflow execution. The
runtime recomputes the threshold for every speculation decision that has
not yet fired. This lets an operator tighten cost controls mid-execution
(e.g., when an SLO budget is being consumed faster than expected) or
relax them (e.g., when an interactive user indicates impatience).

\subsubsection{\texorpdfstring{5.3 \texttt{λ} (time-to-dollars)
specification}{5.3 λ (time-to-dollars) specification}}\label{ux3bb-time-to-dollars-specification}

The latency value ratio \texttt{λ} (USD/second) is a separate input from
\texttt{α}. \texttt{λ} converts latency into monetary value so both
sides of the decision rule live in the same units. Four standard
derivations:

{\def\LTcaptype{none} 
\begin{longtable}[]{@{}
  >{\raggedright\arraybackslash}p{(\linewidth - 4\tabcolsep) * \real{0.3333}}
  >{\raggedright\arraybackslash}p{(\linewidth - 4\tabcolsep) * \real{0.3333}}
  >{\raggedright\arraybackslash}p{(\linewidth - 4\tabcolsep) * \real{0.3333}}@{}}
\toprule\noalign{}
\begin{minipage}[b]{\linewidth}\raggedright
Source
\end{minipage} & \begin{minipage}[b]{\linewidth}\raggedright
Formula
\end{minipage} & \begin{minipage}[b]{\linewidth}\raggedright
Example
\end{minipage} \\
\midrule\noalign{}
\endhead
\bottomrule\noalign{}
\endlastfoot
User value-of-time & operator sets directly & ``1 minute saved = \$1'' →
\texttt{λ\ =\ \$0.0167/s} \\
Labor cost & \texttt{λ\ =\ hourly\_wage\ /\ 3600} & \$100/hr analyst →
\texttt{λ\ =\ \$0.0278/s} \\
Workflow value & \texttt{λ\ =\ value\ /\ expected\_duration} & \$10 per
100-s workflow → \texttt{λ\ =\ \$0.10/s} \\
Budget-deadline & \texttt{λ\ =\ (B\ −\ C₀)\ /\ (T₀\ −\ T)} & derived
from willingness to spend \texttt{B} to hit deadline \texttt{T} \\
\end{longtable}
}

\texttt{α} is a dimensionless preference; \texttt{λ} is a units-bearing
conversion. Keeping them separate is deliberate: \texttt{α} can change
frequently as a UX knob; \texttt{λ} is a deployment-level setting that
rarely changes. Conflating them into one parameter makes cost-preference
updates accidentally re-calibrate the time-to-dollars conversion.

\subsection{\texorpdfstring{6. Dimension 4: EV decision rule with
failure-weighted cost and
\texttt{α}-threshold}{6. Dimension 4: EV decision rule with failure-weighted cost and α-threshold}}\label{dimension-4-ev-decision-rule-with-failure-weighted-cost-and-ux3b1-threshold}

\subsubsection{6.1 The formula}\label{the-formula}

\begin{verbatim}
Latency value:          L_value   = L · λ
Speculation cost:       C_spec    = input_tokens · input_price + output_tokens · output_price
Expected value:         EV        = P · L_value − (1 − P) · C_spec
Threshold:              threshold = (1 − α) · C_spec
Decision:               speculate iff EV ≥ threshold .
\end{verbatim}

On a tie (\texttt{EV\ ==\ threshold}), the default is to speculate:
speculation has potential upside, waiting has none.

\subsubsection{\texorpdfstring{6.2 Failure-weighted cost: why
\texttt{(1−P)}}{6.2 Failure-weighted cost: why (1−P)}}\label{failure-weighted-cost-why-1p}

Several published systems charge speculation cost unconditionally
(B-PASTE's \texttt{µ\ ·\ ΔI}, Speculative Actions v2's \texttt{c\ ·\ m}
in Theorem 4). Our form charges cost only on failure:

\begin{itemize}
\tightlist
\item
  \textbf{Success (probability \texttt{P}).} Speculative result is
  reused; the operation would have been paid either way; the incremental
  cost attributable to \emph{speculation} (as opposed to the work that
  would have happened anyway) is zero. This holds under the
  elastic-capacity assumption of Section 4, where a speculative call
  consumes only its own tokens; under a fixed serving budget there is
  also an opportunity cost on success, treated in Section 14.2.
\item
  \textbf{Failure (probability \texttt{1−P}).} The speculative cost is
  wasted because the operation must be re-executed with the correct
  input. Incremental cost \texttt{=\ C\_spec}.
\end{itemize}

Weighted expectation:
\texttt{EV\ =\ P\ ·\ (+L\_value)\ +\ (1−P)\ ·\ (−C\_spec)\ =\ P\ ·\ L\_value\ −\ (1−P)\ ·\ C\_spec}.

As \texttt{P\ →\ 1}, unconditional-cost formulations diverge materially
from the failure-weighted form. Failure-weighting is the principled form
under pay-per-use API billing, where the cost of \emph{re-execution},
not the cost of running once, is the relevant waste.

\subsubsection{\texorpdfstring{6.3 Why the threshold is
\texttt{(1−α)\ ·\ C\_spec}}{6.3 Why the threshold is (1−α) · C\_spec}}\label{why-the-threshold-is-1ux3b1-c_spec}

Setting \texttt{threshold\ =\ (1−α)\ ·\ C\_spec} gives \texttt{α} a
clean interpretation:

\begin{itemize}
\tightlist
\item
  \texttt{α\ =\ 0} → \texttt{threshold\ =\ C\_spec}: speculate only if
  expected value exceeds the full cost of speculation (i.e., the
  expected gain covers the full worst-case waste).
\item
  \texttt{α\ =\ 1} → \texttt{threshold\ =\ 0}: speculate whenever
  expected value is non-negative.
\item
  Intermediate: linear interpolation.
\end{itemize}

The form also scales with the cost magnitude (\texttt{C\_spec}), so
cheap speculations have a low bar and expensive speculations have a
proportionally higher bar. This is the behavior deployments want: the
same \texttt{α} should produce more-aggressive speculation on cheap
operations and more-conservative speculation on expensive ones.

\subsubsection{6.4 Decision-rule flow}\label{decision-rule-flow}

\pandocbounded{\includegraphics[keepaspectratio]{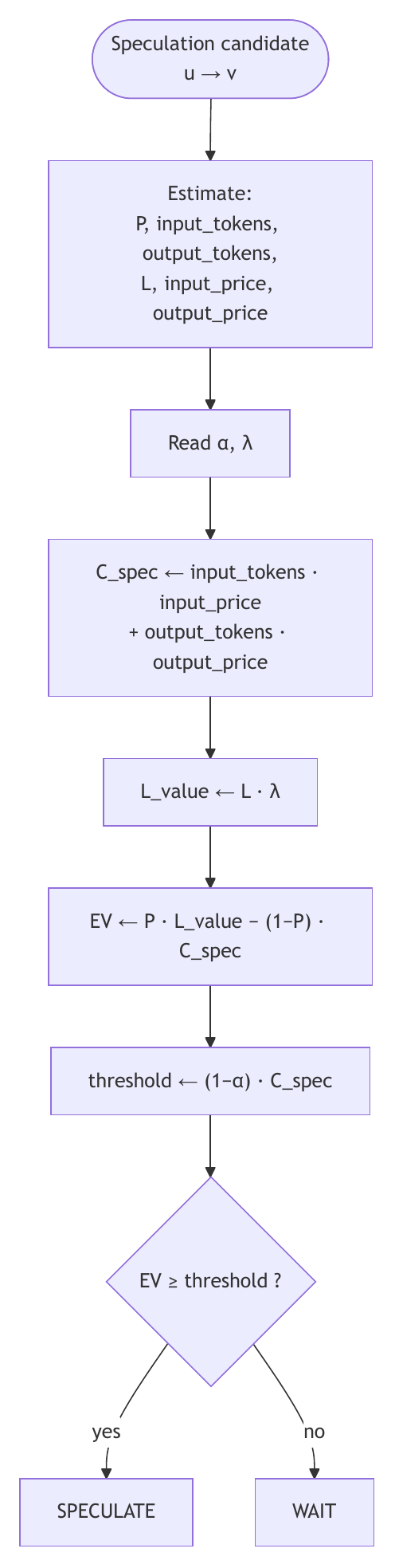}}

\emph{Figure 2. Decision-rule flow. The same formula runs at plan time
and at runtime (Section 8); only the input parameters differ.}

\subsubsection{6.5 Decision-rule
pseudocode}\label{decision-rule-pseudocode}

\begin{Shaded}
\begin{Highlighting}[]
\KeywordTok{def}\NormalTok{ speculation\_decision(}
\NormalTok{    P: }\BuiltInTok{float}\NormalTok{,                       }\CommentTok{\# posterior{-}mean success probability}
\NormalTok{    alpha: }\BuiltInTok{float}\NormalTok{,                   }\CommentTok{\# in [0, 1]}
\NormalTok{    lambda\_dollars\_per\_sec: }\BuiltInTok{float}\NormalTok{,  }\CommentTok{\# deployment{-}configured}
\NormalTok{    input\_tokens: }\BuiltInTok{int}\NormalTok{, output\_tokens: }\BuiltInTok{int}\NormalTok{,}
\NormalTok{    input\_price: }\BuiltInTok{float}\NormalTok{, output\_price: }\BuiltInTok{float}\NormalTok{,}
\NormalTok{    latency\_seconds: }\BuiltInTok{float}\NormalTok{,}
\NormalTok{) }\OperatorTok{{-}\textgreater{}} \BuiltInTok{str}\NormalTok{:}
\NormalTok{    C\_spec    }\OperatorTok{=}\NormalTok{ input\_tokens }\OperatorTok{*}\NormalTok{ input\_price }\OperatorTok{+}\NormalTok{ output\_tokens }\OperatorTok{*}\NormalTok{ output\_price}
\NormalTok{    L\_value   }\OperatorTok{=}\NormalTok{ latency\_seconds }\OperatorTok{*}\NormalTok{ lambda\_dollars\_per\_sec}
\NormalTok{    EV        }\OperatorTok{=}\NormalTok{ P }\OperatorTok{*}\NormalTok{ L\_value }\OperatorTok{{-}}\NormalTok{ (}\DecValTok{1} \OperatorTok{{-}}\NormalTok{ P) }\OperatorTok{*}\NormalTok{ C\_spec}
\NormalTok{    threshold }\OperatorTok{=}\NormalTok{ (}\DecValTok{1} \OperatorTok{{-}}\NormalTok{ alpha) }\OperatorTok{*}\NormalTok{ C\_spec}
    \ControlFlowTok{return} \StringTok{"SPECULATE"} \ControlFlowTok{if}\NormalTok{ EV }\OperatorTok{\textgreater{}=}\NormalTok{ threshold }\ControlFlowTok{else} \StringTok{"WAIT"}
\end{Highlighting}
\end{Shaded}

\subsubsection{\texorpdfstring{6.6 Multi-provider routing under
\texttt{α}}{6.6 Multi-provider routing under α}}\label{multi-provider-routing-under-ux3b1}

An orthogonal consequence: an operation may be routed to different
provider/model tiers based on \texttt{α}. Cost-sensitive preferences
(\texttt{α\ →\ 0}) favor cheaper models; latency-sensitive preferences
(\texttt{α\ →\ 1}) favor faster models. Routing is implemented by
evaluating the decision rule independently per
\texttt{(operation,\ provider,\ model)} candidate and selecting the best
per \texttt{α}. This sits at the boundary of D2 (pricing) and D3
(\texttt{α}) and is a dependent feature of the integrated method.

\subsection{\texorpdfstring{7. Dimension 5: Bayesian \texttt{P} with
structural
priors}{7. Dimension 5: Bayesian P with structural priors}}\label{dimension-5-bayesian-p-with-structural-priors}

Appendix A covers the full posterior-update mechanics and extended
taxonomy entries.

\subsubsection{7.1 The cold-start and small-sample
problem}\label{the-cold-start-and-small-sample-problem}

\texttt{P} is the probability that a speculation is useful; it is the
central quantity in the EV decision rule. Estimating it has three
failure modes:

\begin{enumerate}
\def\labelenumi{\arabic{enumi}.}
\tightlist
\item
  \textbf{Cold start.} When a \texttt{(u,\ v)} pair has never been
  speculatively executed, there is no history to estimate from.
\item
  \textbf{Small-sample regime.} Early in the history, a point estimate
  \texttt{p̂\ =\ successes\ /\ trials} is high variance. A single failure
  in the first two trials gives \texttt{p̂\ =\ 0.5}, which may or may not
  reflect the true rate.
\item
  \textbf{Distribution shift.} Agent behavior changes as prompts are
  edited, models are updated, or upstream operations change their output
  distribution. Old history should be weighted but not dominant.
\end{enumerate}

A Bayesian approach with an informative structural prior and conjugate
updating handles all three.

\subsubsection{7.2 Dependency-type
taxonomy}\label{dependency-type-taxonomy}

The prior on \texttt{P} is selected from a small taxonomy of
\textbf{dependency types}, each capturing a qualitative structural
relationship between upstream output and downstream usability.

{\def\LTcaptype{none} 
\begin{longtable}[]{@{}
  >{\raggedright\arraybackslash}p{(\linewidth - 4\tabcolsep) * \real{0.3056}}
  >{\raggedright\arraybackslash}p{(\linewidth - 4\tabcolsep) * \real{0.1296}}
  >{\raggedright\arraybackslash}p{(\linewidth - 4\tabcolsep) * \real{0.5648}}@{}}
\toprule\noalign{}
\begin{minipage}[b]{\linewidth}\raggedright
Dependency type
\end{minipage} & \begin{minipage}[b]{\linewidth}\raggedright
Prior on \texttt{P}
\end{minipage} & \begin{minipage}[b]{\linewidth}\raggedright
Rationale
\end{minipage} \\
\midrule\noalign{}
\endhead
\bottomrule\noalign{}
\endlastfoot
\texttt{always\_produces\_output} & 0.9 & Upstream always emits;
downstream robust to minor variation. \\
\texttt{list\_output\_variable\_length} & 0.7 & Upstream emits a list;
first-item speculation usually useful. \\
\texttt{conditional\_output} & 0.5 & Upstream output highly variable;
maximum-entropy baseline. \\
\texttt{router\_k\_way} & \texttt{1/k} & Upstream selects 1 of
\texttt{k} downstream paths. \\
\texttt{rare\_event\_trigger} & 0.1--0.2 & Downstream fires only on a
rare upstream signal. \\
\end{longtable}
}

The prior values for \texttt{router\_k\_way} are derived (\texttt{1/k},
not empirical); the values for \texttt{rare\_event\_trigger} are a
narrow range to be pinned per deployment.

\subsubsection{7.3 Beta-Binomial
posterior}\label{beta-binomial-posterior}

\begin{verbatim}
n₀ = 2                            # prior strength (effective sample size)
α₀ = n₀ · p_structural            # Beta shape 1
β₀ = n₀ · (1 − p_structural)      # Beta shape 2

After s successes and f failures on a given (u, v) pair:
    P | data ~ Beta(α₀ + s, β₀ + f)
    E[P | data] = (α₀ + s) / (α₀ + β₀ + s + f) .
\end{verbatim}

Prior mean equals \texttt{p\_structural} by construction. With
\texttt{n₀\ =\ 2}, after roughly 10 observations the posterior mean is
\textasciitilde82\% data-weighted and \textasciitilde18\%
prior-weighted. Appendix A.2 motivates \texttt{n₀\ =\ 2} against larger
alternatives.

\subsubsection{7.4 ``Speculation useful'': three-tier success
criterion}\label{speculation-useful-three-tier-success-criterion}

A speculation with predicted input \texttt{î} and actual input
\texttt{i} is labelled successful if any of:

\begin{itemize}
\tightlist
\item
  \textbf{Tier 1, Exact match:} \texttt{i\ ==\ î}.
\item
  \textbf{Tier 2, Semantic equivalence:} \texttt{equiv(i,\ î)\ ==\ True}
  per a domain predicate. Default: normalized-embedding cosine
  similarity ≥ 0.95 for text; AST equality modulo formatting for code;
  \texttt{semantic\_json\_equal} for structured outputs.
\item
  \textbf{Tier 3, Downstream-output validation} (opt-in, offline): the
  downstream output computed from \texttt{î} is accepted given
  \texttt{i}.
\end{itemize}

Default policy is Tier 1 + Tier 2 with the embedding-similarity
threshold. Tier 3 is opt-in per dependency because it requires running
the actual downstream and comparing results post-hoc, which defeats the
latency benefit on that specific trial (but is fine for offline
calibration).

\subsubsection{7.5 Credible-interval
gating}\label{credible-interval-gating}

The EV decision rule (Section 6) uses a point estimate \texttt{P}. With
a posterior distribution available, the rule can be tightened to the
\texttt{(1−γ)} one-sided lower credible bound:

\begin{verbatim}
P_lower = Beta⁻¹(γ; α₀ + s, β₀ + f)         # e.g., γ = 0.1
speculate iff   P_lower · L_value − (1 − P_lower) · C_spec ≥ (1 − α) · C_spec .
\end{verbatim}

This reduces cold-start over-eagerness: two speculations with identical
posterior means can have very different lower bounds (e.g., 0.80 after
100 trials vs.~0.33 after 2 trials), and only the confident one should
fire.

\subsubsection{\texorpdfstring{7.6 Self-limiting behavior under
branching factor
\texttt{k}}{7.6 Self-limiting behavior under branching factor k}}\label{self-limiting-behavior-under-branching-factor-k}

A natural question about any single-shot speculation scheme is how it
behaves as the upstream's branching factor \texttt{k} grows. The
single-shot rule commits one predicted input \texttt{î} (the mode of the
upstream output distribution) and pays a wasted \texttt{C\_spec}
whenever the actual \texttt{i} falls outside the tier-1/2 equivalence
class of \texttt{î}. Intuition suggests the rule must break down for
large \texttt{k}; the question is \emph{at what k} and \emph{whether the
rule self-limits or silently degrades}.

\textbf{Claim.} The D4 decision rule is self-limiting: as \texttt{k}
grows under a uniform upstream distribution, \texttt{P} falls,
\texttt{EV} collapses below the \texttt{(1−α)\ ·\ C\_spec} threshold,
and the rule correctly produces WAIT without any additional machinery.

\textbf{Critical-k derivation.} Substituting \texttt{P\ =\ 1/k}
(uniform-mode prior, no learned skew) into the D4 rule gives

\begin{verbatim}
EV        = (1/k) · L_value − (1 − 1/k) · C_spec
threshold = (1 − α) · C_spec
\end{verbatim}

Solving \texttt{EV\ ≥\ threshold} for \texttt{k}:

\begin{verbatim}
(1/k) · L_value − (1 − 1/k) · C_spec ≥ (1 − α) · C_spec
(1/k) · (L_value + C_spec)           ≥ (2 − α) · C_spec
k                                    ≤ (L_value + C_spec) / ((2 − α) · C_spec) .
\end{verbatim}

Call this \texttt{k\_crit(α)}. For any
\texttt{k\ \textgreater{}\ k\_crit(α)}, the uniform-case rule WAITs.
Critically, this happens before \texttt{EV} goes negative; the
\texttt{(1−α)·C\_spec} threshold gates off low-but-positive-EV
speculations in proportion to cost-sensitivity.

\textbf{Numerical table at AutoReply parameters.} With
\texttt{L\_value\ =\ \$0.064} and \texttt{C\_spec\ =\ \$0.0135}:

{\def\LTcaptype{none} 
\begin{longtable}[]{@{}llllll@{}}
\toprule\noalign{}
\texttt{k} (uniform) & \texttt{P\ =\ 1/k} & EV & α = 0 & α = 0.5 & α =
1 \\
\midrule\noalign{}
\endhead
\bottomrule\noalign{}
\endlastfoot
2 & 0.500 & +\$0.0253 & SPECULATE & SPECULATE & SPECULATE \\
3 & 0.333 & +\$0.0123 & WAIT & SPECULATE & SPECULATE \\
5 & 0.200 & +\$0.0020 & WAIT & WAIT & SPECULATE \\
10 & 0.100 & −\$0.0058 & WAIT & WAIT & WAIT \\
20 & 0.050 & −\$0.0096 & WAIT & WAIT & WAIT \\
\end{longtable}
}

\texttt{k\_crit(α=1)\ ≈\ 5.7}, \texttt{k\_crit(α=0.5)\ ≈\ 3.8},
\texttt{k\_crit(α=0)\ ≈\ 2.9} at these parameters.

\textbf{Effective k under skewed distributions.} Raw branch count
overstates the adverse case. For a skewed distribution with a dominant
mode of probability \texttt{p\_mode}, the relevant quantity is the
\emph{effective} branching factor \texttt{k\_eff\ =\ 1\ /\ p\_mode},
which may be much smaller than \texttt{k} itself. For example, a 5-way
classifier whose output is 62\% ``billing'' has \texttt{k\_eff\ ≈\ 1.6},
and the EV calculation uses \texttt{P\ =\ 0.62}, not \texttt{P\ =\ 0.2}:

\begin{verbatim}
EV = 0.62 · 0.064 − 0.38 · 0.0135 = +$0.0346 → SPECULATE at all α .
\end{verbatim}

\textbf{D5 bridges the two regimes.} The Bayesian posterior (Section
7.3) begins at the structural prior \texttt{p\ =\ 1/k} (the
\texttt{router\_k\_way} row of the taxonomy, Section 7.2) and converges
to the empirical mode rate as trials accumulate. Cold-start behavior is
governed by Case A (uniform); production steady-state behavior is
governed by Case B (learned). The same mechanism handles both without
case-by-case logic.

\textbf{Remedies when \texttt{k\_eff} is genuinely large.} When the
upstream output distribution is both high-\texttt{k} and flat,
single-shot speculation is not the right abstraction and the EV rule
correctly refuses to fire. Three escape hatches are compatible with the
rest of the method:

\begin{enumerate}
\def\labelenumi{\arabic{enumi}.}
\tightlist
\item
  \textbf{Richer conditioning on \texttt{î}.} Condition the predicted
  input on side-features (e.g., subject-line pattern, time-of-day,
  tenant class) that collapse \texttt{k\_eff} locally. The
  \texttt{router\_k\_way} taxonomy entry of Section 7.2 admits
  per-context variants; a single dependency can host multiple posterior
  cells.
\item
  \textbf{Top-\texttt{m} multi-shot speculation.} Launch
  \texttt{m\ \textgreater{}\ 1} speculations covering the top-\texttt{m}
  modes, paying \texttt{m\ ·\ C\_spec} on all failures but hitting on
  the union of the top-\texttt{m} branches with probability
  \texttt{Σ\_\{i≤m\}\ p\_i}. This is a different decision regime
  (combinatorial over \texttt{m}); B-PASTE's beam admission {[}Song,
  2026, §3{]} and Speculative Actions' integer breadth {[}Ye et al.,
  2025, Thm. 4{]} both live in this space. The single-shot EV rule is a
  degenerate case at \texttt{m\ =\ 1}.
\item
  \textbf{Don't speculate.} The rule already says WAIT. No additional
  guardrail needed.
\end{enumerate}

The self-limiting property means that deploying the method on a workload
with unknown \texttt{k\_eff} is safe: under-estimation of \texttt{P}
leaves money on the table (missed latency wins), but it does not cause
run-away waste.

\subsection{8. Two-phase decision model with bidirectional
override}\label{two-phase-decision-model-with-bidirectional-override}

\subsubsection{8.1 Phase 1: planning}\label{phase-1-planning}

Before execution, the runtime enumerates candidate parallelization plans
and, for each plan, makes a SPECULATE/WAIT decision per candidate
downstream using the Section 6 rule. The planner's objective (combined
with D2/D3/D4) is:

\begin{verbatim}
minimize   α · ( Latency(plan) · λ )  +  (1 − α) · MonetaryCost(plan)
  subject to:
    MonetaryCost(plan) ≤ max_budget        (if specified)
    Latency(plan)      ≤ max_latency       (if specified)
    |wave|             ≤ max_concurrency

MonetaryCost(plan) = Σ_v cost(v)  +  Σ_{spec v} (1 − P_v) · cost_actual(v)
                          base cost              expected speculation waste

Latency(plan)      = Σ_waves  max_{v ∈ wave}  latency(v) .
\end{verbatim}

Candidate plans are generated over discrete concurrency settings
(sequential / maximally parallel / intermediate levels). For small DAGs
(5--20 operations) enumeration is tractable; for larger DAGs,
list-scheduling, ILP, or constraint programming substitute without
changing the rest of the method.

Phase 1 outputs:
\texttt{(plan,\ per-candidate\ SPECULATE/WAIT\ decisions,\ expected\ latency,\ expected\ cost)}.

\subsubsection{8.2 Phase 2: runtime
re-evaluation}\label{phase-2-runtime-re-evaluation}

Immediately before launching any operation marked SPECULATE (and also
before operations marked WAIT that are about to become ready), the
runtime re-runs the Section 6 decision rule with \textbf{current}
parameters:

\begin{itemize}
\tightlist
\item
  Posterior-updated \texttt{P} (D5 / Section 7.3), which may be higher
  or lower than at planning time.
\item
  Updated latency estimates from recent executions (EMA).
\item
  Possibly-changed \texttt{α} (D3 / Section 5.2).
\item
  \texttt{C\_spec} recomputed with current token estimates.
\end{itemize}

The runtime decision can differ from the planning decision in
\textbf{either direction}:

\begin{itemize}
\tightlist
\item
  Planning SPECULATE → Runtime WAIT (downgrade; e.g., \texttt{P} has
  dropped after recent failures).
\item
  Planning WAIT → Runtime SPECULATE (upgrade; e.g., \texttt{α} was
  raised to favor latency).
\end{itemize}

\pandocbounded{\includegraphics[keepaspectratio]{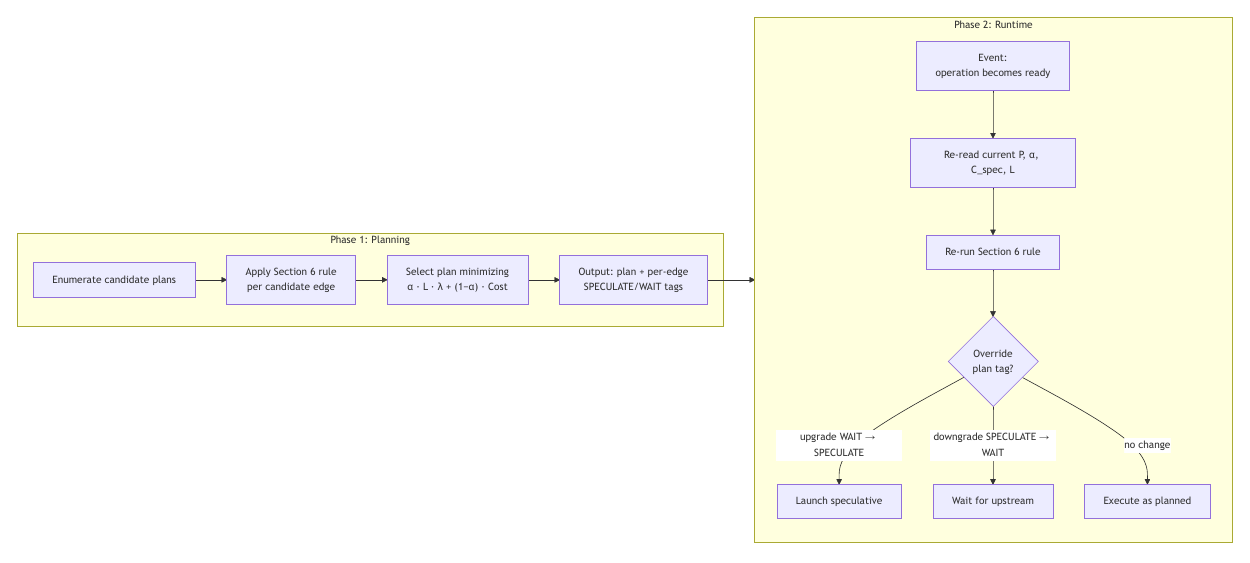}}

\emph{Figure 3. Two-phase decision model. Phase 1 commits a plan and
cost/latency estimate. Phase 2 can override in either direction when
current parameters differ from plan-time parameters.}

\subsubsection{8.3 Why two phases with bidirectional
override}\label{why-two-phases-with-bidirectional-override}

Three reasons:

\begin{enumerate}
\def\labelenumi{\arabic{enumi}.}
\tightlist
\item
  \textbf{Visibility.} Phase 1 produces a cost/latency estimate the user
  can see before execution starts.
\item
  \textbf{Adaptivity.} Phase 2 handles distribution shift, estimate
  drift, and \texttt{α} changes.
\item
  \textbf{Consistency.} Same formula in both phases; only the parameters
  differ.
\end{enumerate}

Bidirectional override matters. One-way ``can only cancel'' is not
enough: a plan-time WAIT may be wrong if the upstream agent has had a
run of successes since planning and its \texttt{P} has risen above the
threshold. Allowing upgrades preserves that option.

\subsection{9. Streaming re-estimation and mid-stream cancellation with
waste
refinement}\label{streaming-re-estimation-and-mid-stream-cancellation-with-waste-refinement}

\subsubsection{\texorpdfstring{9.1 Streaming \texttt{P}
re-estimation}{9.1 Streaming P re-estimation}}\label{streaming-p-re-estimation}

Many modern LLM APIs stream output tokens as they are generated. If the
upstream \texttt{u} streams, the runtime can re-estimate the predicted
input \texttt{î} (and therefore \texttt{P}) as tokens arrive, without
waiting for \texttt{u} to finish.

\begin{verbatim}
At each streamed chunk k of u's output:
    î_k  ← predict_input(partial_output_k)
    P_k  ← P( î_k matches eventual i | u-partial-so-far )
    re-run Section 6 decision rule with P_k .
\end{verbatim}

This gives the speculation decision the benefit of upstream partial
evidence, at the cost of repeated decision-rule evaluation. The decision
rule itself is a handful of multiplies and a comparison (Section 6.5)
and is negligible. The non-trivial per-chunk cost is the input
re-prediction \texttt{predict\_input(partial\_output\_k)} and, at commit
time, the tier-2 equivalence check (Section 7.4), a normalized-embedding
cosine similarity that runs on the critical path. Both must be cheap
relative to the latency being reclaimed for streaming re-estimation to
pay for itself: re-estimation should be throttled (e.g., every
\texttt{N} chunks or on sentence boundaries, not every token) and the
tier-2 embedding model should be small. Section 14.2 (runtime overhead
of the speculation machinery) lists overhead accounting as an open
limitation; a deployment whose tier-2 check or \texttt{î}-predictor is
itself an expensive LLM call may find the machinery costs more latency
than it saves, which the calibration pipeline's offline replay (Section
12.1) will surface as a negative net-latency result before any traffic
is exposed.

\subsubsection{9.2 Mid-stream
cancellation}\label{mid-stream-cancellation}

If at chunk \texttt{k} the updated \texttt{P\_k} falls below the
speculation threshold, the runtime \textbf{cancels} the speculative
downstream mid-execution. Cancellation matters for billing: the
downstream has consumed some fraction of its tokens already.

\subsubsection{\texorpdfstring{9.3 Waste refinement below full
\texttt{C\_spec}}{9.3 Waste refinement below full C\_spec}}\label{waste-refinement-below-full-c_spec}

Naive accounting charges the full \texttt{C\_spec} on cancellation. This
over-estimates waste, because the downstream may have completed only a
fraction \texttt{f\ ∈\ {[}0,1{]}} of its planned generation before
cancellation. Waste refinement tracks the actual token spend:

\begin{verbatim}
C_spec_actual(t) = input_tokens · input_price                      # paid once at launch
                 + tokens_generated_so_far(t) · output_price .

On cancellation at time t:
    C_waste = C_spec_actual(t)    # not full C_spec .
\end{verbatim}

Updating Phase 1's \texttt{Expected\_Speculation\_Waste} to reflect this
reduces the planner's pessimism about speculation:

\begin{verbatim}
Expected_Speculation_Waste_v = (1 − P_v) · E[ C_spec_actual(v) ]
                             = (1 − P_v) · ( C_input + ρ_v · C_output ) ,

where ρ_v ∈ [0,1] is the expected fraction of output generated before cancellation
(estimated from streaming-history EMA; default ρ = 0.5 in the absence of history) .
\end{verbatim}

\pandocbounded{\includegraphics[keepaspectratio]{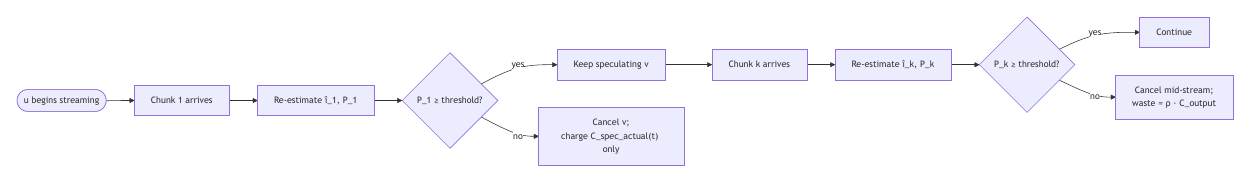}}

\emph{Figure 4. Streaming re-estimation and mid-stream cancellation. On
each new upstream chunk, \texttt{î} is refined, \texttt{P} is
re-estimated, and the decision rule re-evaluated. On cancellation, only
the actually-emitted output tokens are billed, not the full planned
budget.}

The combination of \emph{(streaming re-estimation + mid-stream
cancellation + fractional waste refinement)} is one of the method's
differentiators against the four audited systems (Section 11).

\subsection{10. Worked examples}\label{worked-examples}

\subsubsection{10.1 Single-decision
example}\label{single-decision-example}

\textbf{Setup.}

\begin{itemize}
\tightlist
\item
  Upstream: document-analyzer (producing a list of topics).
\item
  Downstream: topic-researcher.
\item
  Dependency type: \texttt{list\_output\_variable\_length}, prior
  \texttt{p\ =\ 0.7}, 3 successes + 1 failure observed → posterior mean
  \texttt{=\ 4.4\ /\ 6.0\ ≈\ 0.733}.
\item
  Pricing (two-rate example at typical frontier-API prices): input \$3/M
  tokens = \texttt{0.000003\ USD/token}, output \$15/M tokens =
  \texttt{0.000015\ USD/token}.
\item
  Token estimates: input 500, output 1000.
\item
  Estimated latency savings: 5 seconds.
\item
  \texttt{λ\ =\ \$0.01/s}; \texttt{α\ =\ 0.5}.
\end{itemize}

\textbf{Compute.}

\begin{verbatim}
C_spec    = 500 · 0.000003 + 1000 · 0.000015 = 0.0015 + 0.015 = $0.0165
L_value   = 5 · 0.01                         = $0.05
EV        = 0.733 · 0.05 − 0.267 · 0.0165    = 0.03665 − 0.00440 = $0.03225
threshold = (1 − 0.5) · 0.0165               = $0.00825
Decision: EV ($0.03225) ≥ threshold ($0.00825) → SPECULATE .
\end{verbatim}

\textbf{Sensitivity to \texttt{α}} (at \texttt{P\ =\ 0.733}):

{\def\LTcaptype{none} 
\begin{longtable}[]{@{}lll@{}}
\toprule\noalign{}
\texttt{α} & threshold & decision \\
\midrule\noalign{}
\endhead
\bottomrule\noalign{}
\endlastfoot
0 & \$0.01650 & SPECULATE (EV \textgreater{} threshold) \\
0.2 & \$0.01320 & SPECULATE \\
0.5 & \$0.00825 & SPECULATE \\
0.8 & \$0.00330 & SPECULATE \\
1.0 & \$0 & SPECULATE \\
\end{longtable}
}

All \texttt{α} values speculate because EV comfortably exceeds the full
cost. Now lower \texttt{P} to 0.4:

{\def\LTcaptype{none} 
\begin{longtable}[]{@{}llll@{}}
\toprule\noalign{}
\texttt{α} & EV & threshold & decision \\
\midrule\noalign{}
\endhead
\bottomrule\noalign{}
\endlastfoot
0 & \texttt{0.4·0.05\ −\ 0.6·0.0165\ =\ 0.0101} & 0.01650 & WAIT \\
0.2 & 0.0101 & 0.01320 & WAIT \\
0.5 & 0.0101 & 0.00825 & SPECULATE \\
0.8 & 0.0101 & 0.00330 & SPECULATE \\
1.0 & 0.0101 & 0 & SPECULATE \\
\end{longtable}
}

At \texttt{P\ =\ 0.4} the decision flips at \texttt{α\ ≈\ 0.4}. This is
the behavior the \texttt{α}-threshold is designed to produce:
cost-sensitive operators wait, latency-sensitive operators speculate,
and the transition point scales with \texttt{C\_spec}.

\subsubsection{10.2 Two-phase with bidirectional
override}\label{two-phase-with-bidirectional-override}

\textbf{Phase 1 (planning).}

\begin{itemize}
\tightlist
\item
  Same agents as Section 10.1.
\item
  \texttt{P} at planning time: 0.733 (from history).
\item
  Decision: SPECULATE.
\item
  Plan commits to speculative launch.
\end{itemize}

\textbf{Between planning and runtime.}

\begin{itemize}
\tightlist
\item
  Two more trials complete: both failures.
\item
  Posterior updates: \texttt{α\ =\ 4.4,\ β\ =\ 1.6\ +\ 2\ =\ 3.6}, mean
  \texttt{=\ 4.4/8.0\ =\ 0.55}.
\end{itemize}

\textbf{Phase 2 (runtime).}

\begin{itemize}
\tightlist
\item
  Re-evaluate:
  \texttt{EV\ =\ 0.55\ ·\ 0.05\ −\ 0.45\ ·\ 0.0165\ =\ 0.0275\ −\ 0.0074\ =\ \$0.0201}.
\item
  Threshold at \texttt{α\ =\ 0.5}: \$0.00825.
\item
  Still SPECULATE (margin narrowed from \$0.0240 to \$0.0119).
\end{itemize}

\textbf{Alternative: operator raised \texttt{α} to 0.9 between phases.}

\begin{itemize}
\tightlist
\item
  Threshold: \$0.00165. Still SPECULATE (more aggressive, appropriate
  for latency-sensitive preference).
\end{itemize}

\textbf{Alternative: operator lowered \texttt{α} to 0.1 between phases.}

\begin{itemize}
\tightlist
\item
  Threshold: \$0.01485. WAIT.
\item
  Plan said SPECULATE; runtime says WAIT. Bidirectional downgrade in
  action.
\end{itemize}

\subsubsection{10.3 Streaming
cancellation}\label{streaming-cancellation}

\textbf{Setup.} Topic-researcher speculating on an assumed topic
\texttt{î}. Output-generation progress is streamed. Partway through
generation (300 tokens out of estimated 1000 produced), the upstream
completes and produces actual topic \texttt{i}. Tier-1 and Tier-2 checks
on \texttt{i} vs.~\texttt{î} both fail.

\textbf{Waste accounting.}

\begin{verbatim}
C_spec_planned = 500 · 0.000003 + 1000 · 0.000015 = $0.0165
C_spec_actual  = 500 · 0.000003 +  300 · 0.000015 = 0.0015 + 0.0045 = $0.0060
Waste refinement saves:  $0.0165 − $0.0060 = $0.0105  (64% reduction) .
\end{verbatim}

\textbf{Posterior update.} One failure.
\texttt{α\ =\ 4.4,\ β\ =\ 1.6\ +\ 1\ =\ 2.6}, mean \texttt{=\ 0.629}.
Downward revision is appropriate: failures under streaming cancellation
are still real failures for \texttt{P}-estimation purposes, even though
the waste was less than \texttt{C\_spec}.

\subsection{11. Contrast with prior art and
limitations}\label{contrast-with-prior-art-and-limitations}

\subsubsection{11.1 Five-dimension
contrast}\label{five-dimension-contrast}

The four closest audited systems as of May 2026 are:

\begin{itemize}
\tightlist
\item
  \textbf{DSP:} Dynamic Speculative Agent Planning {[}Guan et al.,
  2025{]}.
\item
  \textbf{SA:} Speculative Actions {[}Ye et al., 2025{]}.
\item
  \textbf{SH:} Sherlock {[}Ro et al., 2025{]}.
\item
  \textbf{BP:} B-PASTE {[}Song, 2026{]}.
\end{itemize}

Each cell below is anchored to a specific section or equation in the
cited paper; the compact tags above (DSP, SA, SH, BP) key the table
columns and the per-cell anchors that follow.

{\def\LTcaptype{none} 
\begin{longtable}[]{@{}
  >{\raggedright\arraybackslash}p{(\linewidth - 10\tabcolsep) * \real{0.1667}}
  >{\raggedright\arraybackslash}p{(\linewidth - 10\tabcolsep) * \real{0.1667}}
  >{\raggedright\arraybackslash}p{(\linewidth - 10\tabcolsep) * \real{0.1667}}
  >{\raggedright\arraybackslash}p{(\linewidth - 10\tabcolsep) * \real{0.1667}}
  >{\raggedright\arraybackslash}p{(\linewidth - 10\tabcolsep) * \real{0.1667}}
  >{\raggedright\arraybackslash}p{(\linewidth - 10\tabcolsep) * \real{0.1667}}@{}}
\toprule\noalign{}
\begin{minipage}[b]{\linewidth}\raggedright
Dim
\end{minipage} & \begin{minipage}[b]{\linewidth}\raggedright
DSP
\end{minipage} & \begin{minipage}[b]{\linewidth}\raggedright
Spec Actions v2
\end{minipage} & \begin{minipage}[b]{\linewidth}\raggedright
Sherlock
\end{minipage} & \begin{minipage}[b]{\linewidth}\raggedright
B-PASTE
\end{minipage} & \begin{minipage}[b]{\linewidth}\raggedright
This method
\end{minipage} \\
\midrule\noalign{}
\endhead
\bottomrule\noalign{}
\endlastfoot
D1 & Pre-completion; linear planning chain of \texttt{k} steps {[}DSP
§3, Eq. 1{]} & Pre-response; linear MDP, \texttt{k} future actions {[}SA
Alg. 1, lines 10--15{]} & Post-output, pre-verify; \textbf{structural,
not content-predictive} {[}SH §7 Fig. 9{]} & Reasoning-time,
pattern-guided tool subgraphs {[}BP §3 Eq. 1; Alg. 1{]} &
Pre-completion, arbitrary DAG, LLM-op targets \\
D2 & Dollars in post-hoc eval only; loss uses token counts {[}DSP §5;
Eqs. 6--9{]}¹ & Abstract scalar \texttt{c} (cost-per-unit-time proxy)
{[}SA §5; Thm. 5{]} & \textbf{Single-rate} GPU-hour
\texttt{(unit\_price\ ·\ num\_gpus\ ·\ num\_tokens)\ /\ throughput}
{[}SH App. A.2 Eq. 9{]} & Multi-resource compute profile \texttt{ρ\_i};
no dollars {[}BP Eqs. 1, 4{]} & \textbf{Two-rate} per-token USD (input ≠
output) in the decision rule \\
D3 & \texttt{τ\ ∈\ (0,1)} + \texttt{β} (asymmetric-loss params); 3 named
modes {[}DSP §4.3, §5.3{]} & \texttt{(r,\ c)} offline hyperparameters;
integer \texttt{k} {[}SA §5.2{]} & \textbf{λ on the wrong axis}
(accuracy-vs-cost, not latency-vs-cost) + \texttt{B} (USD cap, not
normalized) {[}SH §6.1 Eq. 1; §7.1 Eq. 5{]} & Multi-resource budget
\texttt{B}; internal \texttt{λ}, \texttt{µ} hyperparameters {[}BP Eqs.
3, 6{]} & \texttt{α\ ∈\ {[}0,1{]}} user-facing, runtime-mutable;
\texttt{λ} (USD/s) separated \\
D4 & TD(λ) value regression with asymmetric loss {[}DSP Eq. 1, §4.1{]};
no \texttt{P}, no cost term in loss &
\texttt{arg\ max\_m\ q(m;\ p)\ ·\ Δ\ −\ c\ ·\ m}; greedy
\texttt{Δ\ ·\ δq\ ≥\ c} {[}SA Thm. 4{]}; cost \textbf{unconditional} &
Hard feasibility:
\texttt{N\_spec\ =\ \{\ j\ :\ Σ\ lat\_exec\ \textless{}\ lat\_vrf\ \}\ ∧\ C\_spec\ ≤\ B}
{[}SH §7.1 Eqs. 4--5{]}; Eq. 6 \texttt{(1−m)\ ·\ ΣC} is closest
structural waste mirror but used for budget check, not decision gate &
\texttt{EU(H\_i)\ =\ q\_i\ ·\ (ΔO\ +\ λ\ ·\ ΔU\ −\ µ\ ·\ ΔI)} {[}BP Eq.
3{]}, \textbf{closest structural EV form}; \texttt{µ\ ·\ ΔI}
unconditional, not \texttt{(1−P)}-weighted; combinatorial over a beam &
\texttt{P\ ·\ L\_value\ −\ (1−P)\ ·\ C\_spec\ ≥\ (1−α)\ ·\ C\_spec};
binary per-candidate gate \\
D5 & DistilBERT regressor, random init, predicts \texttt{k} (not
\texttt{P}) {[}DSP §5, §6.1{]}; paper explicitly rejects a BO baseline &
Model logits or auxiliary classifier; \textbf{constant 0.5 cutoff} {[}SA
§5.2{]} & Empirical match rate \texttt{m\_i} × node-position / fan-in
policy {[}SH §3.3 Fig. 5; §5.4 Alg. 1{]}, \textbf{not} dependency type &
\texttt{q\_i} from offline PrefixSpan frequency counts {[}BP §3{]}; no
runtime Bayesian update & Beta-Binomial posterior, structural prior
keyed to dependency-type taxonomy \\
\end{longtable}
}

\textbf{Notes.}

\begin{enumerate}
\def\labelenumi{\arabic{enumi}.}
\tightlist
\item
  \textbf{DSP abstract vs.~mechanism.} DSP's abstract claims the system
  \emph{``explicitly optimizes a joint objective balancing end-to-end
  latency against dollar cost.''} The TD(λ) training loss {[}DSP Eq.
  1{]} uses token counts, not billing rates, and dollars appear only in
  post-hoc evaluation metrics {[}DSP §5, Eqs. 6--9{]}. The abstract is
  framing; the mechanism does not depend on dollar units.
\item
  \textbf{Sherlock D2 proximity.} The GPU-hour form is linear per-token
  and could be substituted into a decision rule. The surviving D2
  differentiator is two-rate (input ≠ output) API-billing granularity:
  single-rate reductions miss the 3--8× input/output rate asymmetry
  (Section 4.1).
\item
  \textbf{Sherlock D3 axis.} Sherlock's \texttt{λ} trades verifier
  accuracy against cost, not latency against cost. A cost-vs-latency
  dial steers speculation aggressiveness; a cost-vs-accuracy dial steers
  model/verifier tier selection. They are different axes.
\item
  \textbf{B-PASTE D4 credit.} \texttt{EU(H\_i)} is the closest
  structural EV form in the audited corpus. Three differences from this
  method: (a) \texttt{µ\ ·\ ΔI} is unconditional interference, not
  failure-weighted waste; (b) the optimization is combinatorial over an
  admission beam, not a binary gate; (c) it is time-denominated, not
  dollar-denominated.
\end{enumerate}

No paper matches this method on any single dimension, and no subset of
the four matches on the conjunction \emph{(D2 two-rate ∧ D4
failure-weighted + \texttt{(1−α)\ ·\ C} threshold ∧ D5 dependency-type
prior)}.

\subsubsection{11.2 Auxiliary-dimension
contrast}\label{auxiliary-dimension-contrast}

The four neighbors also differ from this method on implementation
choices beyond the five principal dimensions. DSP and Speculative
Actions v2 schedule speculation by tuning \texttt{k}-step or
breadth-\texttt{m} integers; Sherlock places per-node binary
speculations under a topology-driven FSM; B-PASTE runs preemption-first
greedy admission over a subgraph beam. This method runs a per-candidate
binary EV gate inside a two-phase plan-plus-runtime model with
bidirectional override (Section 8). On learning: DSP is online from
random init {[}Guan et al., 2025, §6.1{]}, Speculative Actions is
offline only {[}Ye et al., 2025, §5.1{]}, Sherlock and B-PASTE are
hybrid; this method is online per-\texttt{(u,\ v)} from a structural
cold-start prior. The most informative axis is cancellation behavior:
DSP cancels on upstream-target mismatch {[}Guan et al., 2025, Fig. 1{]}
but does not combine cancellation with streaming re-estimation or
fractional-completion accounting; Speculative Actions discards on cache
miss (correctness, not waste); Sherlock rolls back at node granularity
after the verifier runs; B-PASTE preempts on resource contention, not
input mismatch. The integrated triple (streaming \texttt{P}
re-estimation, mid-stream cancel on input mismatch, and fractional-waste
refinement) appears in no audited system.

\subsubsection{11.3 Adjacent contributions to
credit}\label{adjacent-contributions-to-credit}

Two ideas from B-PASTE are close enough to credit explicitly:

\begin{itemize}
\tightlist
\item
  \textbf{Downstream-unlock-value decomposition}
  \texttt{(ΔO\ +\ λ\ ·\ ΔU)}. The \texttt{L\_value\ =\ L\ ·\ λ} term in
  this paper is end-to-end latency savings and implicitly absorbs unlock
  value for downstream operations on the critical path. For workflows
  where unlock effects propagate multiple hops, the explicit
  decomposition is the right refinement and the decomposed form is a
  strict generalization of what this paper specifies.
\item
  \textbf{Interference term} \texttt{µ\ ·\ ΔI}. In edge or
  shared-compute regimes, co-run interference is a real cost that
  \texttt{(1−P)\ ·\ C\_spec} does not capture (the latter assumes
  pay-per-call API billing). The unified form
  \texttt{EV\ =\ P\ ·\ L\ −\ (1−P)\ ·\ C\ −\ µ\ ·\ ΔI} covers both
  regimes; this paper specifies the API-billing case.
\end{itemize}

Two position papers frame, at the level of general principle, the design
stance this method makes operational:

\begin{itemize}
\tightlist
\item
  \textbf{Bayes-consistent agentic orchestration} {[}Papamarkou et al.,
  2026{]}. Argues that the control layer of an agentic system should be
  Bayes-consistent, maintaining beliefs over task-relevant latent
  quantities and updating them from observed interactions. The position
  is generic: it specifies no posterior, no taxonomy, and no speculation
  mechanism. The method here is one concrete instance of that stance,
  with a Beta-Binomial posterior keyed to the Section 7.2
  dependency-type taxonomy and a closed-form EV gate (Section 6).
\item
  \textbf{Marginal token allocators} {[}Zhu, 2026{]}. Argues that
  agentic systems should be designed as marginal token-allocation
  economies solving
  \texttt{marginal\_benefit\ =\ marginal\_cost\ +\ latency\_cost\ +\ risk\_cost}.
  With \texttt{α} held fixed, this is the same first-order condition as
  the EV rule
  \texttt{EV\ =\ P\ ·\ L\ ·\ λ\ −\ (1−α)\ ·\ C\_spec\ −\ (1−P)\ ·\ C\_spec}.
  That paper is descriptive (no algorithm, no calibration); this method
  is the operational instance, with \texttt{α}-controlled risk
  weighting, a Bayesian estimate of \texttt{P}, and the calibration
  pipeline of Section 12.
\end{itemize}

\subsubsection{11.4 Explicit
differentiators}\label{explicit-differentiators}

Beyond the per-dimension contrast of Section 11.1, five combinations are
individually uncontested across DSP, Speculative Actions v2, Sherlock,
and B-PASTE: (1) a failure-weighted \texttt{(1−P)\ ·\ C} rule paired
with two-rate per-token billing (Sections 4.1, 6); (2) a runtime-mutable
\texttt{α} dial that drives the \texttt{(1−α)\ ·\ C} threshold rather
than the cost (Section 5); (3) a Beta-Binomial posterior keyed to the
dependency-type taxonomy (Section 7.2); (4) two-phase plan-plus-runtime
control with bidirectional override (Section 8); and (5) the integrated
triple of streaming \texttt{P} re-estimation, mid-stream cancellation,
and fractional-waste refinement (Section 9). No neighbor discloses any
of the five, and none discloses their conjunction.

\subsection{12. Calibration and evaluation
pipeline}\label{calibration-and-evaluation-pipeline}

Sections 3--11 specify the decision mechanism; this section specifies
how to tune its free parameters safely and how to decide, per edge,
whether speculation should be enabled at all. Each of the method's
tunable knobs (the dependency-type tag per edge, the prior
\texttt{p\_structural}, the posterior update rate \texttt{n₀}, the
user-facing preference \texttt{α}, the deployment constant \texttt{λ},
the tier-2 equivalence threshold, token estimators, and the per-edge
enable/disable bit) is set or kept honest by one of the five lifecycle
stages below.

The pipeline is staged in order of increasing exposure: offline replay
touches no production traffic, shadow mode serves a decision but
discards it, canary serves a decision to a fraction of traffic, online
calibration runs forever in steady state, and drift detection closes the
loop by flipping the enable bit when any earlier assumption breaks.

\subsubsection{12.1 Offline replay on sequential
logs}\label{offline-replay-on-sequential-logs}

A strictly-sequential deployment of the workflow produces logs of
\texttt{(upstream\_input,\ upstream\_output,\ downstream\_input,\ downstream\_output,\ latency,\ cost)}
tuples. Every calibration step below can be bootstrapped from these logs
before any speculation is enabled.

\textbf{What offline replay computes.}

\begin{itemize}
\tightlist
\item
  \textbf{Effective branching factor.} Fit the empirical upstream-output
  distribution per \texttt{(agent,\ tenant)}. Record \texttt{p\_mode}
  and \texttt{k\_eff\ =\ 1\ /\ p\_mode} (Section 7.6).
\item
  \textbf{Dependency-type auto-assignment.} Assign a type from the
  Section 7.2 taxonomy by rule:
  \texttt{p\_mode\ ≥\ 0.8\ →\ always\_produces\_output}; upstream emits
  a list → \texttt{list\_output\_variable\_length}; \texttt{k\ ≤\ 5}
  with flat distribution → \texttt{router\_k\_way};
  \texttt{p\_mode\ ≤\ 0.2\ →\ rare\_event\_trigger}; otherwise
  \texttt{conditional\_output}. This replaces hand-authored tags for
  edges with sufficient history.
\item
  \textbf{Data-seeded prior.} For each candidate predictor of \texttt{î}
  (modal, regex-extracted, historical-majority, small auxiliary model),
  compute the empirical tier-1 and tier-2 match rate over the logs. Use
  \texttt{(s,\ f)} from the log as \texttt{(s\_0,\ f\_0)} to start the
  posterior, so the edge opens production with \texttt{P} already close
  to truth rather than at the structural prior.
\item
  \textbf{Counterfactual EV grid.} Replay the D4 rule over the logs at a
  grid of \texttt{(α,\ λ)} values and report for each grid point:
  expected latency, expected cost, expected waste, and the realized
  decisions per edge. This is the \textbf{go/no-go per edge} before a
  single dollar of speculative waste is spent.
\end{itemize}

\textbf{What offline replay tunes.} Dependency-type tag; initial P
prior; per-edge enable/disable; deployment-time default \texttt{α}.

\subsubsection{12.2 Shadow mode}\label{shadow-mode}

Launch the speculative downstream alongside the sequential execution,
but commit only the sequential result. Log the full per-decision row
(Appendix C) for every trial.

\textbf{Why it matters.} Shadow mode is the only stage where the runtime
gets posterior data, cost-actual data, tier-2 outcomes, and
fractional-completion evidence without exposing users to any speculative
output. The posterior moves from the cold-start prior to a
data-dominated regime in the same time it would take a live-rolled
speculation to do so, but with zero downside if the tier-2 predicate is
miscalibrated or if the token estimator is badly wrong.

\textbf{What shadow mode tunes.}

\begin{itemize}
\tightlist
\item
  \texttt{P} posterior per edge (to convergence, per Appendix A).
\item
  Tier-2 predicate threshold (default
  \texttt{embedding\_similarity\ ≥\ 0.95} from Section 7.4). Run a grid
  sweep on the shadow logs; select the threshold that maximizes F1
  against a human-graded subset.
\item
  Token estimators (EMA per \texttt{(agent,\ tenant)}; flag
  \texttt{uncertain\_cost} for edges with CoV \textgreater{} threshold).
\item
  Fractional-completion rate \texttt{ρ} for the planner's waste term
  (Section 9.3).
\item
  Credible-bound \texttt{γ} (Section 7.5), if the deployment opts into
  credible-bound gating.
\end{itemize}

Exit criterion: each edge accumulates at least \texttt{N\_shadow} trials
(default 100) with stable posterior mean over the last 50.

\subsubsection{\texorpdfstring{12.3 Canary rollout with \texttt{α} sweep
and implied-\texttt{λ}
recovery}{12.3 Canary rollout with α sweep and implied-λ recovery}}\label{canary-rollout-with-ux3b1-sweep-and-implied-ux3bb-recovery}

Percentage rollout (1\% → 5\% → 25\% → 100\%) of live speculation with a
held-out sequential control. The canary structure:

{\def\LTcaptype{none} 
\begin{longtable}[]{@{}
  >{\raggedright\arraybackslash}p{(\linewidth - 4\tabcolsep) * \real{0.3333}}
  >{\raggedright\arraybackslash}p{(\linewidth - 4\tabcolsep) * \real{0.3333}}
  >{\raggedright\arraybackslash}p{(\linewidth - 4\tabcolsep) * \real{0.3333}}@{}}
\toprule\noalign{}
\begin{minipage}[b]{\linewidth}\raggedright
Arm
\end{minipage} & \begin{minipage}[b]{\linewidth}\raggedright
Primary metric
\end{minipage} & \begin{minipage}[b]{\linewidth}\raggedright
Guardrail metrics
\end{minipage} \\
\midrule\noalign{}
\endhead
\bottomrule\noalign{}
\endlastfoot
Control (sequential) & latency, user-facing CSAT & cost/request, tier-2
false-accept \\
Speculation at \texttt{α\ =\ α\_default} & Δlatency, ΔCSAT, Δcost
vs.~control & posterior drift, waste \$/hr \\
\texttt{α} sweep \texttt{\{0.1,\ 0.3,\ 0.5,\ 0.7,\ 0.9\}} & pareto
(latency, cost) & same \\
\end{longtable}
}

The \texttt{α} sweep traces the empirical \texttt{(latency,\ cost)}
Pareto frontier per edge. The canary is successful when the rollout-rate
arm matches or beats the sequential control on latency while staying
within budget guardrails, and when the \texttt{α} sweep's selected
operating point Pareto-dominates sequential.

\textbf{Implied-\texttt{λ} recovery.} A novel observation enabled by the
\texttt{α}-sweep arm: if operators or automated control loops
consistently prefer an \texttt{α} that the modeled \texttt{λ} would
price out, the modeled \texttt{λ} is under-valued. Concretely, at the
chosen operating point \texttt{α*}, the D4 rule equates

\begin{verbatim}
P · L · λ_implied − (1 − P) · C_spec  =  (1 − α*) · C_spec ,
\end{verbatim}

giving a closed-form recoverable \texttt{λ\_implied} per edge as a
function of observed \texttt{(P,\ C\_spec,\ α*)} and known \texttt{L}.
Averaging over edges and comparing to the deployment's declared
\texttt{λ} yields an audit signal:

\begin{itemize}
\tightlist
\item
  \texttt{λ\_implied} \textgreater{} \texttt{λ\_declared} by a stable
  margin → operators value latency more than the deployment's pricing
  assumes. Refresh \texttt{λ}.
\item
  \texttt{λ\_implied} ≈ \texttt{λ\_declared} → the pricing and
  preferences are consistent.
\item
  \texttt{λ\_implied} \textless{} \texttt{λ\_declared} → the pricing
  over-values latency. Inspect whether the declared \texttt{λ} is based
  on stale CSAT elasticity or stale churn economics.
\end{itemize}

This converts what is otherwise an opaque preference drift into a
dollar-denominated, auditable quantity.

\textbf{What canary tunes.} \texttt{α} operating point per deployment
context (peak vs.~off-hours vs.~incident); \texttt{λ} refresh via the
implied-\texttt{λ} audit; credible-bound \texttt{γ} (final check);
go/no-go to full rollout.

\subsubsection{12.4 Online calibration in steady
state}\label{online-calibration-in-steady-state}

Four continuous checks, ideally on a single dashboard:

\textbf{Posterior calibration curve.} Bucket decisions by predicted
\texttt{P} in width-0.1 buckets. Within each bucket, the empirical
success rate (tier-1 or tier-2 match on the realized \texttt{i}) should
match the bucket midpoint within confidence intervals. Miscalibration
diagnoses:

\begin{itemize}
\tightlist
\item
  Monotonic over-prediction of \texttt{P} → the prior
  \texttt{p\_structural} is too high or the posterior update has missed
  a regime shift. Re-run dependency-type auto-assignment (§12.1) or
  lower \texttt{n₀}.
\item
  Over-confident lower bound → credible-bound \texttt{γ} too low; raise
  it.
\item
  Structural deviation for a specific edge → the dependency-type
  assignment was wrong; re-tag.
\end{itemize}

\textbf{Tier-2 sampling audit.} Sample a fraction (default 1\%) of
committed speculations. Submit to an offline reviewer (human or
LLM-judge) with both \texttt{î} and \texttt{i} available; record whether
the downstream output computed from \texttt{î} would have been
acceptable under \texttt{i} (tier-3 check). False-accept rate above
tolerance (e.g., 5\%) → tighten the tier-2 threshold.

\textbf{Token-estimate monitoring.} Track EMA and CoV of
\texttt{output\_tokens\_actual\ /\ output\_tokens\_est} per
\texttt{(agent,\ tenant)}. High CoV → flag \texttt{uncertain\_cost},
disable speculation on that edge until CoV drops below threshold.

\textbf{\texttt{λ} refresh.} Quarterly, re-run the CSAT-to-churn
regression (or equivalent source-of-truth for latency valuation) on
fresh data; cross-check against the implied-\texttt{λ} audit from §12.3;
update the deployment constant.

\subsubsection{12.5 Drift detection and
kill-switch}\label{drift-detection-and-kill-switch}

Automated triggers that flip the per-edge or global enable bit without
human-in-the-loop approval. This is the mechanism by which
non-stationarity (Section 14.3, non-stationarity response) is handled
operationally.

{\def\LTcaptype{none} 
\begin{longtable}[]{@{}
  >{\raggedright\arraybackslash}p{(\linewidth - 4\tabcolsep) * \real{0.3333}}
  >{\raggedright\arraybackslash}p{(\linewidth - 4\tabcolsep) * \real{0.3333}}
  >{\raggedright\arraybackslash}p{(\linewidth - 4\tabcolsep) * \real{0.3333}}@{}}
\toprule\noalign{}
\begin{minipage}[b]{\linewidth}\raggedright
Trigger
\end{minipage} & \begin{minipage}[b]{\linewidth}\raggedright
Scope
\end{minipage} & \begin{minipage}[b]{\linewidth}\raggedright
Action
\end{minipage} \\
\midrule\noalign{}
\endhead
\bottomrule\noalign{}
\endlastfoot
Posterior mean drops \textgreater{} 20\% over a 100-trial window vs.~the
prior 500 & Edge & Automatically lower \texttt{α\_edge} by 0.2 for the
next hour. \\
Credible lower bound
\texttt{P\_lower\ \textless{}\ (1\ −\ α)\ ·\ C\_spec\ /\ (L\ ·\ λ\ +\ C\_spec)}
for \texttt{N} consecutive decisions & Edge & Disable speculation on the
edge. Require a fresh §12.2 shadow-mode run to re-enable. \\
Tier-2 false-accept rate (§12.4) exceeds tolerance & Edge & Disable
speculation. Page on-call. \\
Monthly cost SLO guardrail tripped & Global & Set \texttt{α\ ←\ 0} for
all edges until next billing cycle. \\
New model version deployed for any agent & All edges using that model &
Flip back to shadow mode for 24 hours; re-run §12.1 auto-assignment on
the shadow logs. \\
Token-estimate CoV \textgreater{} threshold & Edge & Disable speculation
on the edge until CoV drops. \\
\end{longtable}
}

The per-edge enable bit is the method's most consequential operational
knob. Prior sections treat it as given; §12.1 sets it at deployment
time, §12.5 flips it at runtime in response to evidence.

\subsubsection{12.6 Knob-to-stage map}\label{knob-to-stage-map}

{\def\LTcaptype{none} 
\begin{longtable}[]{@{}
  >{\raggedright\arraybackslash}p{(\linewidth - 4\tabcolsep) * \real{0.3333}}
  >{\raggedright\arraybackslash}p{(\linewidth - 4\tabcolsep) * \real{0.3333}}
  >{\raggedright\arraybackslash}p{(\linewidth - 4\tabcolsep) * \real{0.3333}}@{}}
\toprule\noalign{}
\begin{minipage}[b]{\linewidth}\raggedright
Knob
\end{minipage} & \begin{minipage}[b]{\linewidth}\raggedright
Primary stage
\end{minipage} & \begin{minipage}[b]{\linewidth}\raggedright
Secondary
\end{minipage} \\
\midrule\noalign{}
\endhead
\bottomrule\noalign{}
\endlastfoot
Dependency-type tag per edge & §12.1 offline replay & §12.5
drift-triggered re-tag \\
\texttt{p\_structural} prior & §12.1 & §12.2 shadow refinement \\
Posterior \texttt{P} per edge & §12.2 shadow; §12.4 online &
per-decision log \\
\texttt{α} (operating point) & §12.3 canary sweep & post-incident
review \\
\texttt{λ} (time-to-dollars) & §12.3 implied-\texttt{λ} recovery & §12.4
quarterly refresh \\
Tier-2 threshold & §12.2 & §12.4 audit \\
Token estimators & §12.2; §12.4 & §12.5 drift \\
Per-edge enable/disable & §12.1 go/no-go & §12.5 kill-switch \\
Credible-bound \texttt{γ} & §12.3 & §12.4 \\
\end{longtable}
}

Without the full telemetry row of Appendix C, none of the five stages
run.

\subsection{13. Workload fit and archetype
catalog}\label{workload-fit-and-archetype-catalog}

The method targets a specific workload shape. Section 1.4 stated scope
at the abstraction level (static DAG, pay-per-token, predicted
\texttt{î} available); this section restates fit in workload terms
suitable for picking a pilot and for excluding workloads where
single-shot speculation is the wrong tool.

\subsubsection{13.1 Four-point fit rubric}\label{four-point-fit-rubric}

A workload is a good fit for the method when all four of the following
hold:

\begin{enumerate}
\def\labelenumi{\arabic{enumi}.}
\tightlist
\item
  \textbf{Multi-stage workflow with upstream latency to reclaim.} Two or
  more LLM or tool calls with a real upstream wait. Single-call
  workloads have no upstream to speculate against.
\item
  \textbf{Small effective branching factor.} Either raw \texttt{k} is
  low (≤ 5) or the real-world mix is strongly skewed
  (\texttt{p\_mode\ ≥\ 0.5}, equivalently \texttt{k\_eff\ ≤\ 2}). See
  Section 7.6; the method is self-limiting when this fails, but a pilot
  will surface few positive-EV decisions.
\item
  \textbf{Output-heavy downstream.} The speculative downstream generates
  enough tokens for the two-rate pricing (Section 4) to matter. For
  input-heavy, output-light operations, the blended-rate approximation
  error is small and the decision is dominated by latency alone.
\item
  \textbf{Defensible \texttt{λ}.} Someone in the organization can defend
  a USD/second-of-latency-saved figure, typically via CSAT elasticity,
  operator-time value, or an SLA-penalty derivation (Section 5.3). If
  \texttt{λ} cannot be defended, the method collapses to ``always
  speculate'' (\texttt{α\ =\ 1}) or ``never speculate''
  (\texttt{α\ =\ 0}) under the operator's implicit preference.
\end{enumerate}

A workload failing any one of these four is not disqualified, but the
expected yield is low; §12.1 offline replay will show a counterfactual
EV grid dominated by WAIT decisions.

\subsubsection{13.2 Eight archetypes}\label{eight-archetypes}

Production workloads that fit the rubric span four domains. For each
archetype, the entry states the speculation point, the branching
characteristic (raw \texttt{k} or \texttt{k\_eff} under skew), the
business stakes, and the principal watch-out.

\textbf{Customer-facing, real-time.}

\begin{itemize}
\item
  \textbf{Voice bot / IVR response generation.} \emph{Shape:} STT →
  intent classifier → response synthesizer → TTS. \emph{Speculate:}
  response synthesizer with the modal intent's template while the
  classifier runs. \emph{Branching:} per-tenant call mix is Zipfian; top
  3 intents cover 60--80\% (\texttt{k\_eff\ ≈\ 1.5–2}). \emph{Stakes:}
  each additional 400 ms raises call abandonment; telcos pay per minute.
  \emph{Watch-out:} tier-2 equivalence must accept paraphrases (invest
  in the semantic-match predicate).
\item
  \textbf{IDE code autocomplete agent.} \emph{Shape:} context classifier
  (continue-line / complete-block / suggest-test / import-fix) →
  generator. \emph{Speculate:} the generator with the modal intent while
  the classifier inspects surrounding code. \emph{Branching:}
  \textasciitilde70\% of invocations are ``finish this line'' in most
  repos (\texttt{k\_eff\ ≈\ 1.4}). \emph{Stakes:} sub-200 ms feel is the
  product; aggregate GPU hours are real. \emph{Watch-out:}
  cost-sensitivity is high (operators run \texttt{α} near 1 and rely on
  streaming cancellation, §9).
\end{itemize}

\textbf{High-volume enterprise workflows.}

\begin{itemize}
\item
  \textbf{Insurance claims triage.} \emph{Shape:} OCR + claim-type
  classifier → next-action drafter. \emph{Speculate:} drafter for the
  modal next-action per claim type. \emph{Branching:} 3--4 claim
  archetypes cover most volume per insurer (\texttt{k\_eff\ ≈\ 2–3}).
  \emph{Stakes:} adjuster time at \$50--100/hr; 20\% cycle-time
  reduction scales to seven-figure annual savings. \emph{Watch-out:}
  tier-3 offline validation is mandatory (regulatory risk);
  credible-bound gating (Section 7.5) from day one.
\item
  \textbf{Content-moderation triage.} \emph{Shape:} safety classifier →
  action drafter (allow / warn / remove / escalate). \emph{Speculate:}
  the ``allow'' path with its user-facing message. \emph{Branching:}
  extreme skew; ``allow'' wins at \texttt{p\_mode\ \textgreater{}\ 0.95}
  (\texttt{k\_eff\ ≈\ 1.05}). \emph{Stakes:} platforms process billions
  of items/day; unit wins compound. \emph{Watch-out:} the rare non-allow
  paths are where quality matters most; tier-2 must never be softened
  for them.
\item
  \textbf{Medical prior-authorization drafting.} \emph{Shape:} document
  extraction → procedure-code classifier → policy retrieval →
  approval/denial drafter. \emph{Speculate:} the retrieval + drafter
  path for the modal code. \emph{Branching:} per payer-specialty, top
  5--10 codes cover most volume (\texttt{k\_eff\ ≈\ 3–5}).
  \emph{Stakes:} prior-auth backlogs delay hospital revenue; each day
  shaved is directly monetizable. \emph{Watch-out:} cold-start on new
  payers is high-risk (wrong denial draft); credible-bound gating from
  day one plus shadow-mode runway per new payer.
\end{itemize}

\textbf{Developer-tooling workflows.}

\begin{itemize}
\item
  \textbf{PR / code-review bot.} \emph{Shape:} diff analyzer →
  change-type classifier (feature / bugfix / refactor / config / docs) →
  review-strategy selector → reviewer prompt. \emph{Speculate:} the
  reviewer prompt for the modal change type per repo. \emph{Branching:}
  per-repo skew is strong (frontend repos \textasciitilde60\% features,
  infra repos \textasciitilde50\% config), so \texttt{k\_eff\ ≈\ 2} per
  repo even with \texttt{k\ =\ 5} raw. \emph{Stakes:} reviewer wait time
  is engineering velocity; at organization scale this is a
  multi-million-dollar lever. \emph{Watch-out:} cross-repo
  generalization is weak; rely on per-repo posteriors (the method does
  this by default).
\item
  \textbf{RAG query-answering pipeline.} \emph{Shape:} intent classifier
  (factual / multi-hop / comparison / procedural) → retriever strategy →
  answer synthesizer. \emph{Speculate:} the synthesizer with the
  most-likely intent's retrieval path. \emph{Branching:} factual lookup
  dominates at 60--70\% for most products (\texttt{k\_eff\ ≈\ 1.5–2}).
  \emph{Stakes:} user-facing latency drives engagement; output-heavy
  synthesis is the expensive stage. \emph{Watch-out:} the retriever is
  itself a tool call and may be slow; consider speculation at the
  retriever level separately.
\end{itemize}

\textbf{High-stakes, low-volume.}

\begin{itemize}
\tightlist
\item
  \textbf{Security alert / incident triage.} \emph{Shape:} alert
  enricher → alert-type classifier → runbook selector → remediation-plan
  drafter. \emph{Speculate:} remediation drafter for the most-likely
  runbook. \emph{Branching:} top 5 alert types dominate most SOCs;
  time-of-incident tightens further (\texttt{k\_eff\ ≈\ 2–3}).
  \emph{Stakes:} MTTR has dollar value in breach exposure;
  incident-minutes of operator time are expensive. \emph{Watch-out:} low
  volume per unique alert → posterior converges slowly; lean on the
  structural prior (\texttt{router\_k\_way}) longer than in a
  high-volume setting.
\end{itemize}

\subsubsection{13.3 Where the method does not
fit}\label{where-the-method-does-not-fit}

Four workload shapes where single-shot speculation is the wrong tool and
no amount of tuning helps:

\begin{itemize}
\tightlist
\item
  \textbf{Open-ended creative generation} (prompt → long-form essay, one
  call). The downstream \emph{is} the workflow; there is no upstream to
  speculate against.
\item
  \textbf{Runtime-determined topology} (reflection loops, dynamic
  spawning, recursive planning). Section 1.4 scopes these out. Each
  expansion requires re-planning, and the planner assumptions (Section
  8.1) do not hold.
\item
  \textbf{High \texttt{k\_eff} with flat distribution.} Single-shot
  speculation's EV collapses below threshold (Section 7.6); remedies are
  richer conditioning on \texttt{î}, top-\texttt{m} multi-shot
  speculation, or declining to speculate.
\item
  \textbf{Cheap-downstream workloads.} When \texttt{C\_spec} and
  \texttt{L\ ·\ λ} are both small, EV is small by construction and
  rarely clears the \texttt{(1\ −\ α)\ ·\ C\_spec} threshold. The
  decision rule correctly says WAIT, but the effort of instrumentation
  has no payoff.
\end{itemize}

\subsubsection{13.4 Pilot-picking rubric}\label{pilot-picking-rubric}

Score candidate workloads on:

\begin{enumerate}
\def\labelenumi{\arabic{enumi}.}
\tightlist
\item
  Is there latency pain a user or operator actually feels?
\item
  Is there a single upstream-output mode hitting above 50\%
  (\texttt{k\_eff\ ≤\ 2})?
\item
  Is the downstream output-heavy (so D2's two-rate pricing moves the
  decision)?
\item
  Can \texttt{P} be instrumented observationally (§12.1 offline replay,
  §12.2 shadow) before turning speculation on?
\end{enumerate}

Workloads scoring high on all four (voice-bot, claims triage,
moderation, code-review bot among §13.2) are the best first pilots.
Workloads that need streaming cancellation (§9) to be economical (IDE
autocomplete, RAG) are second-tier. Workloads that need credible-bound
gating from day one (prior-auth, security triage) require a longer
shadow-mode runway.

\subsection{14. Limitations and open
problems}\label{limitations-and-open-problems}

The limitations below fall into two kinds: those addressed operationally
by the calibration pipeline (Sections 12--13) but lacking a closed-form
solution, and those that are genuinely open with no mitigation yet. Each
item states which. They are grouped into three categories --- scope
boundaries, cost-model completeness, and estimation and calibration ---
with the cost-model items placed first because they bear most directly
on the central expected-value claim.

\subsubsection{14.1 Scope boundaries}\label{scope-boundaries}

\begin{itemize}
\tightlist
\item
  \textbf{Static DAG assumption.} The planner (Section 8.1) assumes a
  fixed topology. Dynamic workflows (loops, reflection, dynamic
  spawning) would require treating each runtime expansion as its own
  planning sub-problem against the Section 8.1 rule; this extension is
  open and unaddressed.
\item
  \textbf{Streaming availability.} Section 9's waste refinement requires
  APIs that stream output tokens \textbf{and} support mid-stream
  cancellation. Most major providers support both as of 2026; some
  tool-call APIs do not, in which case the method falls back to
  full-\texttt{C\_spec} waste accounting (Section 6) with no loss of
  correctness, only of the streaming saving.
\end{itemize}

\subsubsection{14.2 Cost-model
completeness}\label{cost-model-completeness}

These two items bear directly on the EV accounting and are the most
consequential limitations in this section.

\begin{itemize}
\tightlist
\item
  \textbf{Capacity contention and opportunity cost.} The cost model
  (Section 4) assumes pay-per-call API billing with effectively elastic
  capacity: a speculative call's only cost is its own tokens. Under a
  fixed serving budget (a rate-limit ceiling, a reserved GPU fleet, or a
  self-hosted endpoint at high utilization), this assumption fails. A
  speculation then consumes capacity that would otherwise serve live,
  non-speculative requests, so an aggressive \texttt{α} can \emph{raise}
  tail latency for the rest of the workload even as it lowers latency
  for the speculated edge. The \texttt{(1−P)\ ·\ C\_spec} term prices
  the wasted tokens but not this opportunity cost. The
  interference-augmented EV form credited to B-PASTE in Section 11.3 is
  the right hook for the contended-capacity regime; this paper specifies
  only the elastic-API case. A deployment near its capacity ceiling
  should treat speculation as drawing from a shared budget and gate it
  accordingly (the §12.5 global cost-SLO trigger is a coarse version of
  this); a principled per-decision opportunity-cost term is open.
\item
  \textbf{Runtime overhead of the speculation machinery.} The method's
  own bookkeeping is not free, and a latency-reclaiming method must show
  its overhead is dominated by the latency it reclaims. The EV
  evaluation is negligible (Section 6.5), but three components are not:
  per-chunk input re-prediction under streaming (Section 9.1), the
  tier-2 embedding-similarity check on the critical path at commit time
  (Section 7.4), and the posterior read/update per decision (Section
  7.3). When the \texttt{î}-predictor or the tier-2 check is itself a
  non-trivial model call, the machinery can cost more latency than it
  saves. The method mitigates this by throttling re-estimation and
  recommending small tier-2 models (Section 9.1), and the offline-replay
  stage (Section 12.1) flags a net-negative-latency edge before any
  traffic is exposed --- but flagging is not measuring. A measured
  overhead characterization on a live deployment, and a closed-form
  overhead budget folded into the EV rule rather than caught empirically
  downstream, are both open; Appendix D is synthetic and does not
  measure machinery overhead.
\end{itemize}

\subsubsection{14.3 Estimation and
calibration}\label{estimation-and-calibration}

\begin{itemize}
\tightlist
\item
  \textbf{Token-estimation variance.} Sensitivity to \texttt{C\_spec}
  error rises as \texttt{P} decreases. Agents with high output-length
  variance should be tagged \texttt{uncertain\_cost} per §12.2 and §12.5
  and excluded until history stabilizes (operationally mitigated).
\item
  \textbf{\texttt{λ} elicitation and the two-parameter interface.} The
  four methods in Section 5.3 cover most cases, but genuinely
  cost-insensitive settings (e.g., research exploration) are better
  served by \texttt{α\ =\ 1} than by a large \texttt{λ}; §12.3's
  implied-\texttt{λ} recovery is an audit, not an independent source of
  truth. Exposing both \texttt{α} (preference) and \texttt{λ}
  (conversion) is also a UX burden: a single-dial interface is
  implementable by fixing \texttt{λ} from deployment config and
  surfacing only \texttt{α}, but the underlying method requires both.
\item
  \textbf{Prior-value provenance.} The specific values 0.9 / 0.7 / 0.5
  in Section 7.2 are design choices motivated by the semantic taxonomy.
  §12.1's data-seeded prior replaces the taxonomy default for any edge
  with sufficient history; large-scale empirical calibration across
  public workflow corpora nonetheless remains open.
\item
  \textbf{Non-stationarity response.} The default prior strength
  \texttt{n₀\ =\ 2} favors responsiveness over stability. §12.5 drift
  triggers handle non-stationarity operationally (by flipping the enable
  bit or lowering \texttt{α}); a closed-form decayed-observation scheme
  that down-weights older trials (e.g., exponential forgetting / a
  discounted Beta update) is a natural complement rather than a
  replacement, and is open.
\item
  \textbf{Joint estimation.} Two dependencies of the same type in the
  same workflow may share information; a hierarchical Bayesian model
  could pool evidence. Each \texttt{(u,\ v)} pair currently gets an
  independent belief (open).
\item
  \textbf{Tier-2 threshold calibration.} The default
  \texttt{embedding\_similarity\ ≥\ 0.95} is a reasonable starting
  value. §12.2 and §12.4 sampled audits provide per-deployment tuning;
  cross-domain benchmarks are open.
\end{itemize}

\subsection{References}\label{references}

\begin{itemize}
\tightlist
\item
  {[}Guan et al., 2025{]} Guan, Y., Lan, Q., Sun, F., Ding, D., Acharya,
  D., Wang, C., Wang, W. Y., Hua, W. \emph{Dynamic Speculative Agent
  Planning} (DSP). arXiv:2509.01920 {[}cs.AI{]}, 2025.
\item
  {[}Leviathan et al., 2023{]} Leviathan, Y., Kalman, M., Matias, Y.
  \emph{Fast Inference from Transformers via Speculative Decoding.}
  arXiv:2211.17192 {[}cs.LG{]}, 2022.
\item
  {[}Papamarkou et al., 2026{]} Papamarkou, T., Alquier, P., Bauer, M.,
  Buntine, W., Davison, A., Dziugaite, G. K., Filippone, M., Foong, A.
  Y. K., Fortuin, V., Fouskakis, D., Frellsen, J., Hüllermeier, E.,
  Karaletsos, T., Khan, M. E., Kotelevskii, N., Lahlou, S., Li, Y., Liu,
  F., Lyle, C., Möllenhoff, T., Palla, K., Panov, M., Sale, Y.,
  Schweighofer, K., Shelmanov, A., Swaroop, S., Trapp, M., Waegeman, W.,
  Wilson, A. G., Zaytsev, A. \emph{Position: Agentic AI Orchestration
  Should Be Bayes-Consistent.} arXiv:2605.00742 {[}cs.AI{]}, 2026.
\item
  {[}Ro et al., 2025{]} Ro, Y., Qiu, H., Goiri, Í., Fonseca, R.,
  Bianchini, R., Akella, A., Wang, Z., Erez, M., Choukse, E.
  \emph{Sherlock: Reliable and Efficient Agentic Workflow Execution.}
  arXiv:2511.00330 {[}cs.MA{]}, 2025.
\item
  {[}Song, 2026{]} Song, Y. \emph{B-PASTE: Beam-Aware Pattern-Guided
  Speculative Execution for Resource-Constrained LLM Agents.}
  arXiv:2604.16469 {[}cs.DC{]}, 2026.
\item
  {[}Sui et al., 2026{]} Sui, Y., Zhao, H., Ma, R., He, Z., Wang, H.,
  Li, J., Yang, Y. \emph{Act While Thinking: Accelerating LLM Agents via
  Pattern-Aware Speculative Tool Execution} (PASTE). arXiv:2603.18897
  {[}cs.AI{]}, 2026.
\item
  {[}Ye et al., 2025{]} Ye, N., Ahuja, A., Liargkovas, G., Lu, Y.,
  Kaffes, K., Peng, T. \emph{Speculative Actions: A Lossless Framework
  for Faster Agentic Systems.} arXiv:2510.04371 {[}cs.AI{]}, 2025.
\item
  {[}Zhu, 2026{]} Zhu, S. \emph{Agentic AI Systems Should Be Designed as
  Marginal Token Allocators.} arXiv:2605.01214 {[}cs.AI{]}, 2026.
\end{itemize}

\subsection{Appendix A: Bayesian posterior
mechanics}\label{appendix-a-bayesian-posterior-mechanics}

\subsubsection{A.1 Conjugate pair}\label{a.1-conjugate-pair}

\texttt{P} is modeled as a random variable in \texttt{{[}0,1{]}}. The
prior is Beta; each speculation outcome is a Bernoulli trial (success =
``speculation useful'' per Section 7.4); by conjugacy the posterior is
Beta.

\begin{verbatim}
Prior:        P ~ Beta(α₀, β₀)
Observation:  for each trial i ∈ {1..n}, X_i ~ Bernoulli(P)
              s = Σ X_i  (successes),  f = n − s  (failures)
Posterior:    P | data ~ Beta(α₀ + s, β₀ + f) .
\end{verbatim}

\subsubsection{\texorpdfstring{A.2 Why \texttt{n₀\ =\ 2} and not larger
or
smaller}{A.2 Why n₀ = 2 and not larger or smaller}}\label{a.2-why-nux2080-2-and-not-larger-or-smaller}

A degenerate prior with \texttt{α₀\ =\ β₀\ =\ 0} (improper Beta) gives
the maximum-likelihood estimate directly after one observation. Two
reasons not to use it:

\begin{enumerate}
\def\labelenumi{\arabic{enumi}.}
\tightlist
\item
  With \texttt{n₀\ =\ 0}, a single failure on trial 1 gives posterior
  mean 0, which collapses the decision rule to ``never speculate again''
  for that dependency, since a single noisy outcome has disproportionate
  influence.
\item
  The structural prior carries real information (the expected average
  success rate for this dependency type). Discarding it forfeits the
  benefit of the taxonomy.
\end{enumerate}

At the other extreme, \texttt{n₀\ =\ 100} gives a very stiff prior; it
takes \textasciitilde100 observations to move the posterior mean
materially. Given the non-stationarity of LLM-agent workflows (model
updates, prompt edits), responsiveness to recent evidence matters more
than stability. \texttt{n₀\ =\ 2} is the smallest integer that retains
the structural prior as a tie-breaker without overwhelming early
observations.

\subsubsection{A.3 Verification table}\label{a.3-verification-table}

With \texttt{n₀\ =\ 2} and \texttt{α₀\ +\ β₀\ =\ 2} in every row below:

{\def\LTcaptype{none} 
\begin{longtable}[]{@{}
  >{\raggedright\arraybackslash}p{(\linewidth - 8\tabcolsep) * \real{0.3438}}
  >{\raggedright\arraybackslash}p{(\linewidth - 8\tabcolsep) * \real{0.1250}}
  >{\raggedright\arraybackslash}p{(\linewidth - 8\tabcolsep) * \real{0.1667}}
  >{\raggedright\arraybackslash}p{(\linewidth - 8\tabcolsep) * \real{0.1250}}
  >{\raggedright\arraybackslash}p{(\linewidth - 8\tabcolsep) * \real{0.2396}}@{}}
\toprule\noalign{}
\begin{minipage}[b]{\linewidth}\raggedright
Dependency type
\end{minipage} & \begin{minipage}[b]{\linewidth}\raggedright
\texttt{p\_struct}
\end{minipage} & \begin{minipage}[b]{\linewidth}\raggedright
\texttt{(α₀,\ β₀)}
\end{minipage} & \begin{minipage}[b]{\linewidth}\raggedright
Prior mean
\end{minipage} & \begin{minipage}[b]{\linewidth}\raggedright
Prior mode
\end{minipage} \\
\midrule\noalign{}
\endhead
\bottomrule\noalign{}
\endlastfoot
\texttt{always\_produces\_output} & 0.9 & (1.8, 0.2) & 0.900 & 1.0
(boundary) \\
\texttt{list\_output\_variable\_length} & 0.7 & (1.4, 0.6) & 0.700 & 1.0
(boundary) \\
\texttt{conditional\_output} & 0.5 & (1.0, 1.0) & 0.500 & undefined
(uniform) \\
\texttt{router\_k\_way} (k=3) & 0.333 & (0.667, 1.333) & 0.333 & 0.0
(boundary) \\
\end{longtable}
}

When \texttt{α₀\ \textless{}\ 1} or \texttt{β₀\ \textless{}\ 1}, the
Beta density goes to infinity at the corresponding boundary. This is
statistically correct (the prior is pulling toward that boundary) and is
harmless for the decision rule: the posterior mean, not the mode, drives
decisions, and after one interior observation the density is bounded.

\subsubsection{A.4 Posterior update worked
example}\label{a.4-posterior-update-worked-example}

Agent A (document analyzer) → Agent B (topic researcher). Dependency
tagged \texttt{list\_output\_variable\_length} (\texttt{p\ =\ 0.7},
\texttt{n₀\ =\ 2}, so \texttt{α₀\ =\ 1.4,\ β₀\ =\ 0.6}).

{\def\LTcaptype{none} 
\begin{longtable}[]{@{}lllllll@{}}
\toprule\noalign{}
Step & Event & Successes & Failures & \texttt{α} & \texttt{β} &
Posterior mean \\
\midrule\noalign{}
\endhead
\bottomrule\noalign{}
\endlastfoot
0 & Initial & 0 & 0 & 1.4 & 0.6 & 0.700 \\
1 & Success & 1 & 0 & 2.4 & 0.6 & 0.800 \\
2 & Success & 2 & 0 & 3.4 & 0.6 & 0.850 \\
3 & Failure & 2 & 1 & 3.4 & 1.6 & 0.680 \\
4 & Success & 3 & 1 & 4.4 & 1.6 & 0.733 \\
5--10 & Mixed: 5 successes, 0 failures & 8 & 1 & 9.4 & 1.6 & 0.855 \\
\end{longtable}
}

After 10 observations the posterior mean has converged to 0.855,
weighted \textasciitilde82\% by data (8/9 empirical success) and
\textasciitilde18\% by prior (0.700). The intended balance:
responsiveness with a sanity-check prior.

\subsubsection{A.5 Credible-bound example: cold-start
vs.~mature}\label{a.5-credible-bound-example-cold-start-vs.-mature}

Two dependencies, both with posterior mean 0.85:

{\def\LTcaptype{none} 
\begin{longtable}[]{@{}lllll@{}}
\toprule\noalign{}
Scenario & \texttt{α} & \texttt{β} & Mean & 10\% lower bound \\
\midrule\noalign{}
\endhead
\bottomrule\noalign{}
\endlastfoot
Mature & 85 & 15 & 0.850 & 0.803 \\
Cold-start & 1.7 & 0.3 & 0.850 & 0.325 \\
\end{longtable}
}

The mature dependency is confidently \textasciitilde0.85; the cold-start
dependency (1 success after an \texttt{always\_produces\_output} prior)
has wide uncertainty. Under \texttt{P\_min\ =\ 0.5}:

\begin{itemize}
\tightlist
\item
  Mean-based rule: both speculate.
\item
  Lower-bound rule: only the mature one speculates; the cold-start one
  waits for more data.
\end{itemize}

\subsection{Appendix B: Router-dependency
example}\label{appendix-b-router-dependency-example}

Agent A routes to exactly one of three downstream agents (B, C, D).
Speculation candidate: Agent B. Dependency tagged
\texttt{router\_k\_way} with \texttt{k=3} (prior \texttt{p\ =\ 1/3},
\texttt{α₀\ =\ 2/3,\ β₀\ =\ 4/3}).

{\def\LTcaptype{none} 
\begin{longtable}[]{@{}
  >{\raggedright\arraybackslash}p{(\linewidth - 10\tabcolsep) * \real{0.1667}}
  >{\raggedright\arraybackslash}p{(\linewidth - 10\tabcolsep) * \real{0.1667}}
  >{\raggedright\arraybackslash}p{(\linewidth - 10\tabcolsep) * \real{0.1667}}
  >{\raggedright\arraybackslash}p{(\linewidth - 10\tabcolsep) * \real{0.1667}}
  >{\raggedright\arraybackslash}p{(\linewidth - 10\tabcolsep) * \real{0.1667}}
  >{\raggedright\arraybackslash}p{(\linewidth - 10\tabcolsep) * \real{0.1667}}@{}}
\toprule\noalign{}
\begin{minipage}[b]{\linewidth}\raggedright
Trial
\end{minipage} & \begin{minipage}[b]{\linewidth}\raggedright
Actual route
\end{minipage} & \begin{minipage}[b]{\linewidth}\raggedright
Speculation of B useful?
\end{minipage} & \begin{minipage}[b]{\linewidth}\raggedright
\texttt{α}
\end{minipage} & \begin{minipage}[b]{\linewidth}\raggedright
\texttt{β}
\end{minipage} & \begin{minipage}[b]{\linewidth}\raggedright
Posterior mean
\end{minipage} \\
\midrule\noalign{}
\endhead
\bottomrule\noalign{}
\endlastfoot
0 & n/a & n/a & 0.667 & 1.333 & 0.333 \\
1 & B & Success & 1.667 & 1.333 & 0.556 \\
2 & C & Failure & 1.667 & 2.333 & 0.417 \\
3 & B & Success & 2.667 & 2.333 & 0.533 \\
4 & D & Failure & 2.667 & 3.333 & 0.444 \\
5 & B & Success & 3.667 & 3.333 & 0.524 \\
\end{longtable}
}

After 5 trials with 3 successes (60\% empirical) and the 33\% prior, the
posterior mean is 0.524, data-weighted toward the empirical rate but
anchored by the prior. If the router has genuine skew (e.g., B is
selected 60\% of the time), the posterior will converge to roughly 0.6
with more data; if the router is uniform over \texttt{k=3}, the
posterior will regress toward 0.33.

\subsection{Appendix C: Telemetry
schema}\label{appendix-c-telemetry-schema}

Every calibration and evaluation stage of Section 12 consumes the same
per-decision log row. Without it, none of the stages run. This appendix
specifies the schema.

\subsubsection{C.1 Per-decision row}\label{c.1-per-decision-row}

\begin{Shaded}
\begin{Highlighting}[]
\ImportTok{from}\NormalTok{ dataclasses }\ImportTok{import}\NormalTok{ dataclass}
\ImportTok{from}\NormalTok{ typing }\ImportTok{import}\NormalTok{ Literal, Optional}

\AttributeTok{@dataclass}
\KeywordTok{class}\NormalTok{ SpeculationDecision:}
    \CommentTok{\# identity}
\NormalTok{    decision\_id: }\BuiltInTok{str}                       \CommentTok{\# UUID, unique per candidate edge event}
\NormalTok{    trace\_id: }\BuiltInTok{str}                          \CommentTok{\# workflow execution id (joins across decisions)}
\NormalTok{    edge: }\BuiltInTok{tuple}\NormalTok{[}\BuiltInTok{str}\NormalTok{, }\BuiltInTok{str}\NormalTok{]                  }\CommentTok{\# (upstream\_agent, downstream\_agent)}
\NormalTok{    dep\_type: Literal[}
        \StringTok{"always\_produces\_output"}\NormalTok{,}
        \StringTok{"list\_output\_variable\_length"}\NormalTok{,}
        \StringTok{"conditional\_output"}\NormalTok{,}
        \StringTok{"router\_k\_way"}\NormalTok{,}
        \StringTok{"rare\_event\_trigger"}\NormalTok{,}
\NormalTok{    ]}
\NormalTok{    tenant: }\BuiltInTok{str}                            \CommentTok{\# per{-}tenant posteriors require this key}
\NormalTok{    model\_version: }\BuiltInTok{tuple}\NormalTok{[}\BuiltInTok{str}\NormalTok{, }\BuiltInTok{str}\NormalTok{]         }\CommentTok{\# (agent, version) for drift{-}triggered re{-}tag}

    \CommentTok{\# decision inputs (at evaluation time)}
\NormalTok{    alpha: }\BuiltInTok{float}                           \CommentTok{\# in [0, 1]}
\NormalTok{    lambda\_usd\_per\_s: }\BuiltInTok{float}
\NormalTok{    P\_mean: }\BuiltInTok{float}                          \CommentTok{\# Beta posterior mean}
\NormalTok{    P\_lower\_bound: Optional[}\BuiltInTok{float}\NormalTok{]         }\CommentTok{\# γ{-}credible lower bound, if gating}
\NormalTok{    C\_spec\_est\_usd: }\BuiltInTok{float}
\NormalTok{    L\_est\_s: }\BuiltInTok{float}                         \CommentTok{\# estimated latency savings on success}
\NormalTok{    input\_tokens\_est: }\BuiltInTok{int}
\NormalTok{    output\_tokens\_est: }\BuiltInTok{int}
\NormalTok{    input\_price: }\BuiltInTok{float}                     \CommentTok{\# USD/token}
\NormalTok{    output\_price: }\BuiltInTok{float}                    \CommentTok{\# USD/token}

    \CommentTok{\# decision outputs}
\NormalTok{    EV\_usd: }\BuiltInTok{float}
\NormalTok{    threshold\_usd: }\BuiltInTok{float}
\NormalTok{    decision: Literal[}\StringTok{"SPECULATE"}\NormalTok{, }\StringTok{"WAIT"}\NormalTok{]}
\NormalTok{    phase: Literal[}\StringTok{"plan"}\NormalTok{, }\StringTok{"runtime"}\NormalTok{]      }\CommentTok{\# Section 8 two{-}phase model}
\NormalTok{    overrode: Literal[}\StringTok{"none"}\NormalTok{, }\StringTok{"upgrade"}\NormalTok{, }\StringTok{"downgrade"}\NormalTok{]  }\CommentTok{\# runtime vs. plan}
\NormalTok{    i\_hat\_source: Literal[}
        \StringTok{"modal"}\NormalTok{, }\StringTok{"regex"}\NormalTok{, }\StringTok{"historical"}\NormalTok{, }\StringTok{"stream\_k"}\NormalTok{, }\StringTok{"auxiliary\_model"}
\NormalTok{    ]}

    \CommentTok{\# guardrails / audit (set at decision time)}
\NormalTok{    uncertain\_cost\_flag: }\BuiltInTok{bool}              \CommentTok{\# set by §12.4 EMA monitor}
\NormalTok{    enabled: }\BuiltInTok{bool}                          \CommentTok{\# §12.5 kill{-}switch state at decision time}
\NormalTok{    budget\_remaining\_usd: Optional[}\BuiltInTok{float}\NormalTok{]  }\CommentTok{\# for cost SLO triggers}

    \CommentTok{\# realized outcomes (filled in after upstream completes; default None)}
\NormalTok{    i\_actual: Optional[}\BuiltInTok{object}\NormalTok{] }\OperatorTok{=} \VariableTok{None}      \CommentTok{\# full upstream output for replay}
\NormalTok{    tier1\_match: Optional[}\BuiltInTok{bool}\NormalTok{] }\OperatorTok{=} \VariableTok{None}
\NormalTok{    tier2\_match: Optional[}\BuiltInTok{bool}\NormalTok{] }\OperatorTok{=} \VariableTok{None}
\NormalTok{    tier3\_accept: Optional[}\BuiltInTok{bool}\NormalTok{] }\OperatorTok{=} \VariableTok{None}    \CommentTok{\# filled offline, sampled (§12.4)}
\NormalTok{    committed\_speculative: }\BuiltInTok{bool} \OperatorTok{=} \VariableTok{False}    \CommentTok{\# True → kept; False → re{-}ran with i}
\NormalTok{    C\_spec\_actual\_usd: Optional[}\BuiltInTok{float}\NormalTok{] }\OperatorTok{=} \VariableTok{None}  \CommentTok{\# Section 9.3 fractional waste}
\NormalTok{    tokens\_generated\_before\_cancel: Optional[}\BuiltInTok{int}\NormalTok{] }\OperatorTok{=} \VariableTok{None}
\NormalTok{    latency\_actual\_s: Optional[}\BuiltInTok{float}\NormalTok{] }\OperatorTok{=} \VariableTok{None}
\end{Highlighting}
\end{Shaded}

The dataclass above is a complete specification; realized-outcome fields
default to \texttt{None} (or \texttt{False} for the boolean) so that a
row can be emitted at decision time and filled in later when the
upstream operation completes.

\subsubsection{C.2 Deriving every calibration signal from one
row}\label{c.2-deriving-every-calibration-signal-from-one-row}

{\def\LTcaptype{none} 
\begin{longtable}[]{@{}
  >{\raggedright\arraybackslash}p{(\linewidth - 2\tabcolsep) * \real{0.5000}}
  >{\raggedright\arraybackslash}p{(\linewidth - 2\tabcolsep) * \real{0.5000}}@{}}
\toprule\noalign{}
\begin{minipage}[b]{\linewidth}\raggedright
Signal
\end{minipage} & \begin{minipage}[b]{\linewidth}\raggedright
Derivation
\end{minipage} \\
\midrule\noalign{}
\endhead
\bottomrule\noalign{}
\endlastfoot
Posterior update (§7.3) &
\texttt{(s,\ f)\ ←\ (s,\ f)\ +\ (tier1\_match\ ∨\ tier2\_match,\ ¬(tier1\_match\ ∨\ tier2\_match))}
per edge \\
Effective \texttt{k} (§7.6) & empirical distribution of
\texttt{i\_actual} per \texttt{(edge,\ tenant)} \\
Counterfactual EV grid (§12.1) & replay D4 over the row with varied
\texttt{(α,\ λ)} \\
Tier-2 false-accept rate (§12.4) & fraction of
\texttt{committed\_speculative\ ∧\ ¬tier3\_accept} over sampled rows \\
Token-estimate CoV (§12.4) &
\texttt{std(tokens\_generated\_before\_cancel\ /\ output\_tokens\_est)}
over committed rows; on full-completion rows,
\texttt{tokens\_generated\_before\_cancel} equals the actual output
count \\
Implied-\texttt{λ} (§12.3) & solve
\texttt{P\ ·\ L\ ·\ λ\ −\ (1−P)\ ·\ C\_spec\ =\ (1−α)\ ·\ C\_spec} for
\texttt{λ} at observed \texttt{α*} \\
Waste per failed speculation (§9.3) & \texttt{C\_spec\_actual\_usd} when
\texttt{¬committed\_speculative} \\
Cost SLO burn & \texttt{Σ\ C\_spec\_actual\_usd} over budget window \\
Drift trigger (§12.5) & posterior-mean delta over rolling windows per
edge \\
\end{longtable}
}

\subsubsection{C.3 Retention and sampling
policy}\label{c.3-retention-and-sampling-policy}

Per-decision rows are small (\textless{} 1 KB serialized). A reasonable
default policy:

\begin{itemize}
\tightlist
\item
  Retain \emph{all} rows for 30 days (on-call debugging, drift
  attribution).
\item
  Retain \emph{aggregated} \texttt{(edge,\ tenant,\ day)} posterior
  updates indefinitely (the Bayesian state).
\item
  Retain \emph{sampled} rows (1\%) indefinitely for long-term
  calibration studies.
\item
  \texttt{tier3\_accept} is filled offline from the sampled pool; humans
  or LLM-judges consume a working queue populated from those samples.
\end{itemize}

The schema is intentionally flat. Join keys
\texttt{(decision\_id,\ trace\_id,\ edge,\ tenant,\ model\_version)}
allow cross-cutting queries without requiring a specific data-warehouse
shape.

\subsection{Appendix D: Synthetic numerical
validation}\label{appendix-d-synthetic-numerical-validation}

This appendix validates the method's closed-form behavior at the
canonical AutoReply parameters (Section 10) through five seeded
synthetic experiments. Each experiment is fully specified below as a
direct evaluation of an equation stated in Sections 4--9, driven only by
\texttt{numpy} under a single fixed seed (\texttt{seed\ =\ 20260531});
the reported numbers are stable across runs and reproducible from the
specification alone.

These experiments validate the \emph{decision rule and its supporting
mechanisms} against their own equations. They are \textbf{not}
measurements of any deployed LLM workload. The calibration pipeline in
Section 12 specifies how to extend this validation to a real deployment.

No code, data, or other files accompany this submission. Each experiment
is reconstructible from the equation it evaluates and the AutoReply
parameters, under the single fixed seed above for the synthetic
Bernoulli draws; no proprietary dependency, dataset, or LLM call is
involved.

\subsubsection{D.1 Decision-boundary
validation}\label{d.1-decision-boundary-validation}

The grid
\texttt{(k,\ α)\ ∈\ \{1,\ …,\ 10\}\ ×\ \{0,\ 0.25,\ 0.5,\ 0.75,\ 1.0\}}
is swept at AutoReply parameters with \texttt{P\ =\ 1/k} (Section 7.6
Case A, uniform prior over \texttt{k} branches). Each cell applies the
EV decision rule and records SPECULATE or WAIT. The closed-form
critical-\texttt{k} curve is overlaid in red.

\pandocbounded{\includegraphics[keepaspectratio]{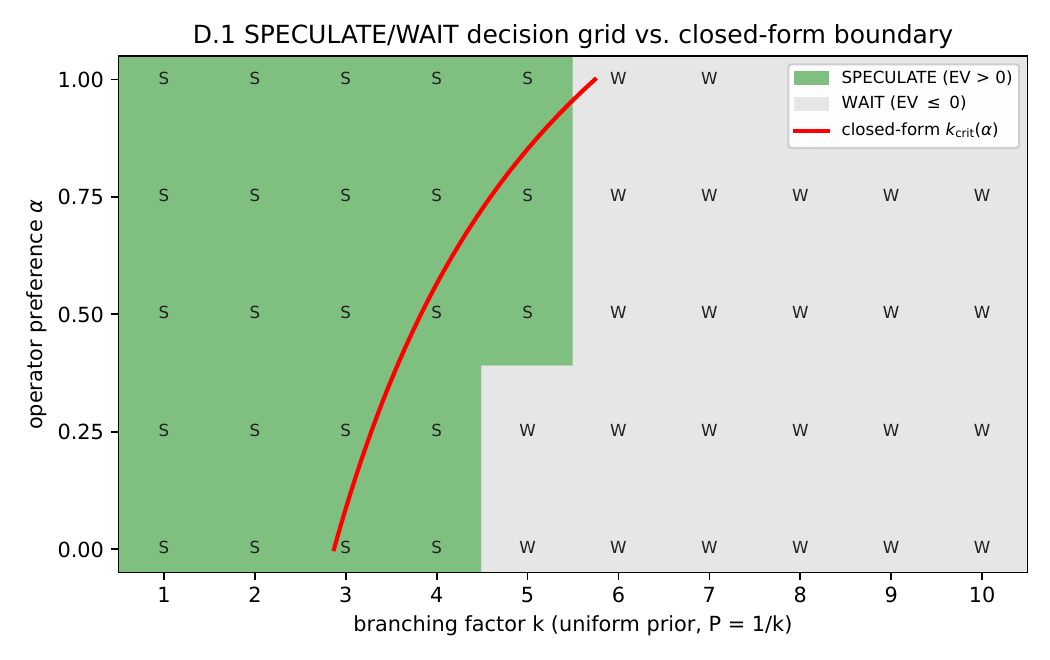}}

The empirical SPECULATE/WAIT boundary lies exactly along
\texttt{k\_crit(α)\ =\ (L\_value\ +\ C\_spec)\ /\ ((2\ -\ α)\ ·\ C\_spec)}.
At AutoReply parameters this gives \texttt{k\_crit(0)} approximately
2.87, \texttt{k\_crit(0.5)} approximately 3.83, and
\texttt{k\_crit(1.0)} approximately 5.74, matching Section 7.6's table.
The validation confirms that the rule self-limits as the upstream
branching factor grows: under uniform prior, no \texttt{α} in {[}0, 1{]}
makes the rule SPECULATE for \texttt{k} at least 6 at this scenario's
economics.

\subsubsection{D.2 P-threshold
validation}\label{d.2-p-threshold-validation}

With \texttt{α\ =\ 0.5} fixed at AutoReply parameters, \texttt{P} is
swept over \texttt{{[}0.05,\ 0.95{]}} and EV is plotted. The closed-form
break-even is \texttt{P*\ =\ C\_spec\ /\ (L\_value\ +\ α\ ·\ C\_spec)}.

\pandocbounded{\includegraphics[keepaspectratio]{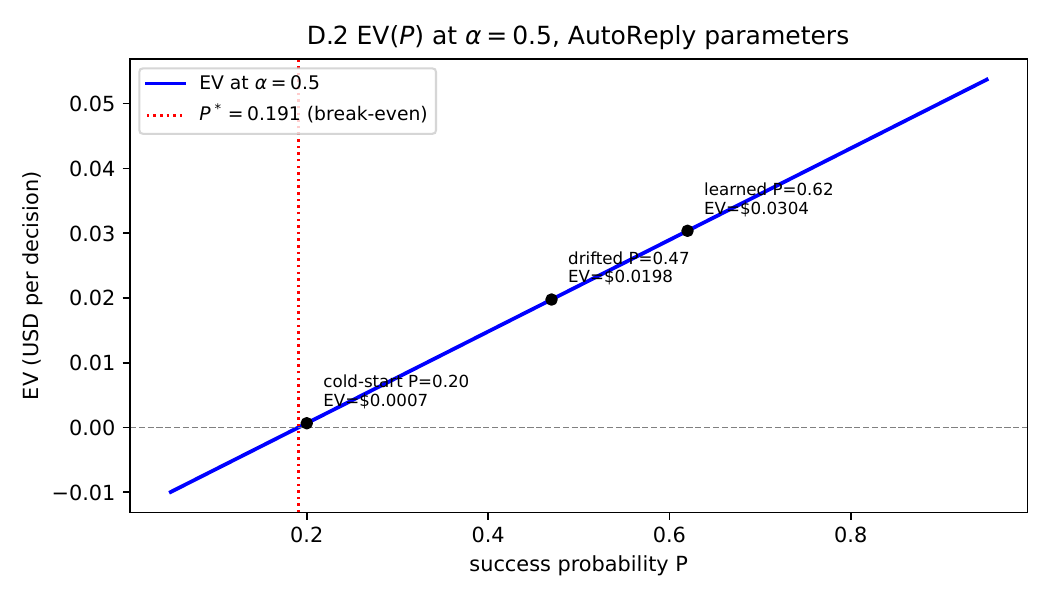}}

At AutoReply parameters, \texttt{P*} is approximately 0.19 for
\texttt{α\ =\ 0.5}. The cold-start uniform prior \texttt{P\ =\ 0.20}
sits just above break-even (EV approximately +\$0.0007, borderline
SPECULATE), the post-drift posterior \texttt{P\ =\ 0.47} clears it
comfortably (EV approximately +\$0.020), and the learned steady state
\texttt{P\ =\ 0.62} is well above (EV approximately +\$0.030). The same
EV rule produces qualitatively different decisions across the three
regimes without any parameter tuning, exactly the Section 10 worked
example.

\subsubsection{D.3 Bayesian posterior
convergence}\label{d.3-bayesian-posterior-convergence}

Starting from the \texttt{conditional\_output} structural prior
\texttt{Beta(1,\ 1)} (Section 7.2, \texttt{n₀\ =\ 2}), 200 sequential
observations are drawn from a Bernoulli with \texttt{P\_true\ =\ 0.62}
and the Beta-Binomial posterior is updated. Posterior mean and 95\%
credible interval are tracked over time.

\pandocbounded{\includegraphics[keepaspectratio]{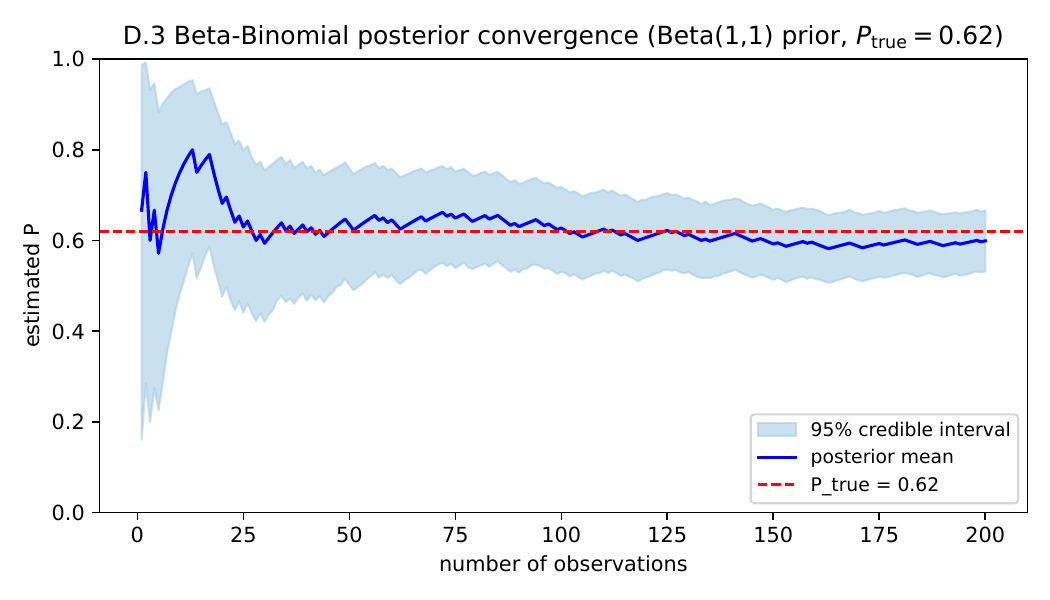}}

The posterior mean enters the neighborhood of \texttt{P\_true} within
roughly 30 observations and the 95\% credible interval narrows
continuously. After 200 observations the interval is
\texttt{{[}0.53,\ 0.67{]}} (per-seed), consistent with the standard
\texttt{1/√n} shrinkage from Beta posteriors. This validates the §7
claim that production traffic supplies sufficient signal to recover the
operative \texttt{P} without tuned hyperparameters.

\subsubsection{D.4 Streaming cancellation waste
reduction}\label{d.4-streaming-cancellation-waste-reduction}

The validation simulates 10,000 speculative attempts at AutoReply
parameters with \texttt{P\_success\ =\ 0.62}. Failed speculations are
aborted mid-stream after a fraction \texttt{f} of Agent B's output
tokens have been emitted, paying \texttt{C\_input\ +\ f\ ·\ C\_output}
instead of the full \texttt{C\_spec}. Three policies are compared: no
streaming (every failure costs full \texttt{C\_spec}), mean-cancel at
\texttt{f\ =\ 0.37}, and per-attempt random cancel
\texttt{f\ ∼\ Unif{[}0.10,\ 0.60{]}}.

\pandocbounded{\includegraphics[keepaspectratio]{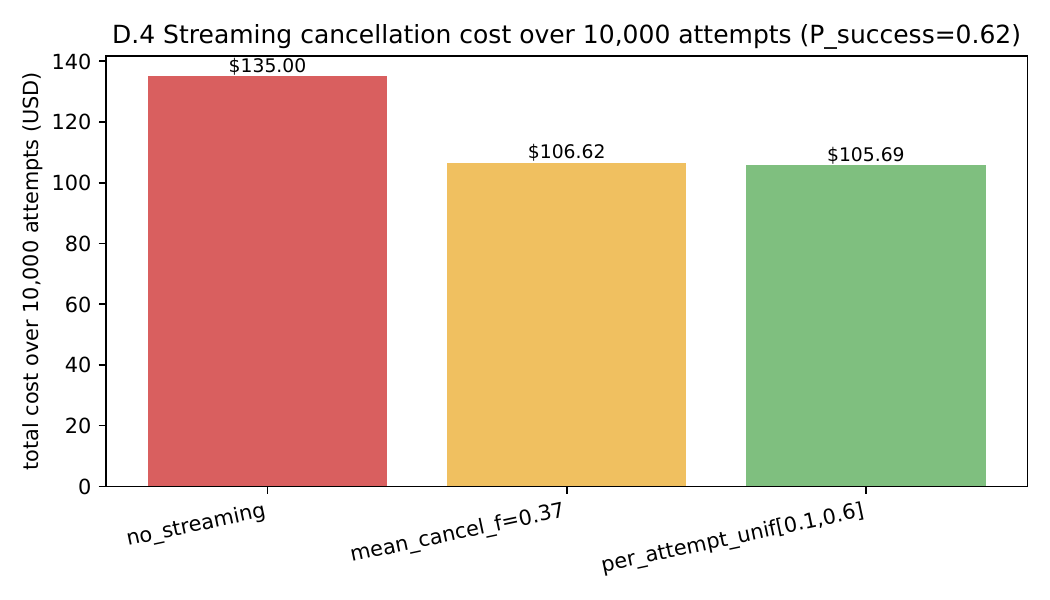}}

Mean-cancel reduces total cost from \$135.00 to \$106.62 (a 21\% saving)
with per-failure waste dropping from \$0.0135 to \$0.0059 (a 56\% drop,
matching the headline number in Section 9). The per-attempt random
cancel produces a similar saving (\$105.69) at this distribution of
cancel fractions. The §9.3 fractional-waste accounting is therefore not
a marginal optimization: it is the difference between losing the full
speculative cost on every failure and recovering most of it.

\textbf{Telemetry-schema conformance.} This evaluation is structured so
that each simulated decision carries the full field set of the Appendix
C \texttt{SpeculationDecision} schema (33 fields), and the aggregate
plot and cost summary are derived only from those per-decision records,
demonstrating, on synthetic data, the same discipline §C.2 claims is
sufficient to reconstruct every calibration signal in a real deployment.
The other four experiments exercise their closed-form equations directly
over the quantities each plot needs; a real-deployment harness would
record the full schema for every decision regardless of which signal it
serves.

\subsubsection{\texorpdfstring{D.5 Implied-\texttt{λ} recovery (audit
signal)}{D.5 Implied-λ recovery (audit signal)}}\label{d.5-implied-ux3bb-recovery-audit-signal}

For each \texttt{α*\ ∈\ {[}0,\ 1{]}}, the EV equation is solved
backwards for the \texttt{λ} that would make \texttt{α*} the rational
operating point at AutoReply \texttt{(P,\ C\_spec,\ L\_upstream)}:

\texttt{λ\_implied\ =\ {[}(1\ −\ α*)\ ·\ C\_spec\ +\ (1\ −\ P)\ ·\ C\_spec{]}\ /\ (P\ ·\ L\_upstream)}.

This is the §12.3 audit signal: if operators settle at an \texttt{α*}
whose implied \texttt{λ} is far from the declared \texttt{λ},
calibration must refresh.

\pandocbounded{\includegraphics[keepaspectratio]{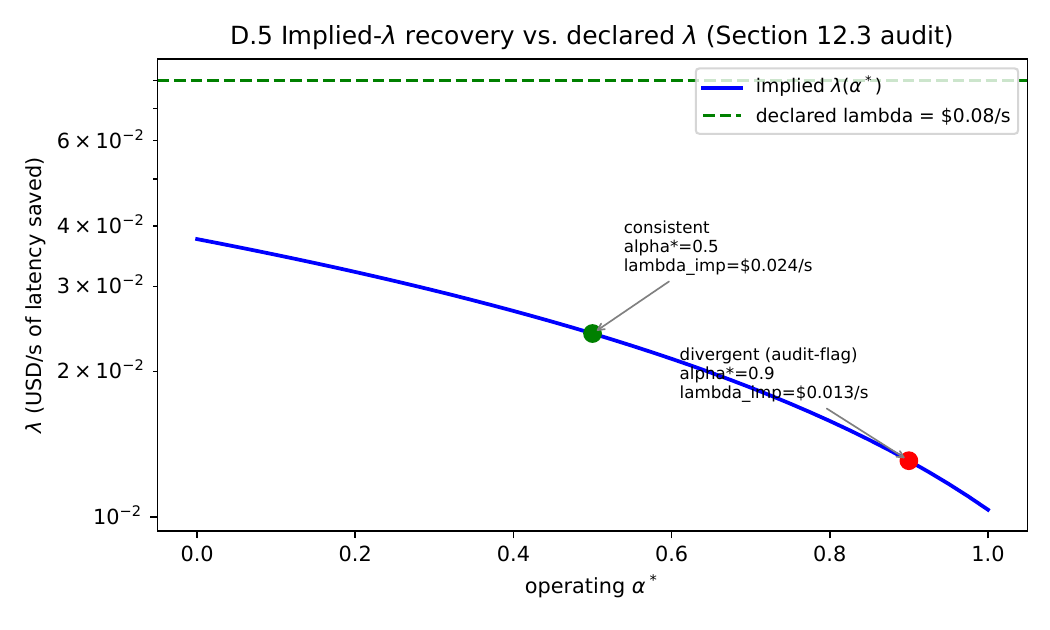}}

At \texttt{α*\ =\ 0.5} the implied \texttt{λ} is approximately
\$0.024/s, broadly comparable to the declared \$0.08/s, consistent with
operators behaving as the latency-dollars conversion suggests. At
\texttt{α*\ =\ 0.9} (the paper's running example) the implied \texttt{λ}
collapses to roughly \$0.013/s, an order of magnitude below the declared
value. Operators driving the dial that aggressively are revealing a far
lower marginal value of latency than the deployment-configured
\texttt{λ} claims, the audit-flag scenario in Section 12.3.

\subsubsection{D.6 Threats to validity}\label{d.6-threats-to-validity}

These experiments validate the method against its own equations and
assumptions at the AutoReply parameters. They do not establish that:

\begin{itemize}
\tightlist
\item
  Real LLM workloads have stationary \texttt{P}. The calibration
  pipeline in Section 12 exists precisely because \texttt{P} drifts; D.5
  is the audit signal.
\item
  Latency and token cost are deterministic point estimates. Real
  deployments see distributions; the streaming re-estimation in Section
  9 is the runtime mechanism for this.
\item
  The AutoReply parameters generalize. Section 13's workload-fit rubric
  is the way to evaluate a new scenario.
\end{itemize}

A real-deployment evaluation following Section 12.1 → 12.5 (offline
replay → shadow → canary with \texttt{α} sweep → online calibration →
drift-triggered kill-switch) is the natural next step and the explicit
goal of the calibration pipeline.

\end{document}